\documentclass[11pt]{article}

\usepackage{comment}
\usepackage[pdftex]{color, graphicx}
\usepackage{hyperref} 
\usepackage{graphics}
\usepackage{subcaption}
\usepackage{epsfig}
\usepackage{latexsym}
\usepackage{mathrsfs}
\usepackage{eufrak}
\usepackage{amsmath}
\usepackage{graphicx}
\usepackage{longtable}
\usepackage{url}
\usepackage{authblk}
\usepackage{multicol}
\usepackage{fancyvrb}  
\usepackage{lmodern}   
\usepackage{changepage}

\newcommand{\dt}{\ensuremath{\delta t}}
\newcommand{\ff}[2]{\frac{#1}{#2}}
\newcommand{\pp}[2]{\frac{\partial #1}{\partial #2}}
\newcommand{\ddt}[1]{\frac{\partial #1}{\partial t}}

\renewcommand{\vec}[1]{\ensuremath{\mathbf{#1}}}
\newcommand{\tensor}[1]{\mathsf{#1}}

\newcommand{\B}{\vec{B}}
\newcommand{\E}{\vec{E}}
\newcommand{\J}{\vec{J}}

\renewcommand{\v}{\vec{v}}

\renewcommand{\P}{\tensor{\Pi}}

\newcommand{\grad}[1]{\nabla #1}

\renewcommand{\div}[1]{\nabla \cdot #1}

\newcommand{\curl}[1]{\nabla \times #1}

\usepackage{graphicx}
\usepackage{dcolumn}
\usepackage{bm}

\begin{document}


\title{Fast solvers for tokamak fluid models with PETSc}%

\author[1]{Mark F. Adams} 
\author[2]{Jin Chen}%
\author[3]{Benjamin Sturdevant}
\affil[1]{Lawrence Berkeley National Laboratory, Berkeley CA}%
\affil[2]{Princeton Plasma Physics Laboratory, Princeton NJ}%
\affil[3]{Rensselaer Polytechnic Institute, SCOREC, Troy NY}%

\date{}

\maketitle

\begin{abstract}

Multigrid (MG) is widely recognized as a highly effective solver for the model problem, the Laplacian, but textbook MG fails on most problems of interest. 
MG methods have been applied to complex, real-world applications with careful consideration of the physical model and discretization.
This work develops the first step in applying MG methods to science and engineering relevant magnetohydrodynamics (MHD) tokamak models in the \textit{M3D-C1}\footnote{\url{https://m3dc1.pppl.gov}} fusion energy science code. 
The semi-implicit time integrator in \textit{M3D-C1} is composed of many linear solves.
The implicit advance of the momentum equation is the most challenging and is the focus of this work.

The current production solver in \textit{M3D-C1} is a block Jacobi (BJ) preconditioner within a Krylov solver, where blocks group degrees of freedom on planes of constant toroidal coordinate. BJ convergence degrades as the number of planes increases due to the spectral properties of the matrix preconditioned with BJ.
The partially magnetic field-aligned, regular toroidal grid structure in \textit{M3D-C1} is amenable to semi-coarsening geometric MG in the toroidal direction.
This paper develops such a solver and demonstrates competitive performance on a runaway electron model of a SPARC\footnote{\url{https://cfs.energy/technology/sparc}} disruption, and superior robustness on a stellarator model on which the BJ solver fails to converge.

\end{abstract}



\section{\label{sec:level1}Introduction}

\textit{M3D-C1} is a code that solves the extended-magnetohydrodynamic (MHD) equations, which is a model that describes plasma as electrically conducting fluids of ions and electrons. This code is primarily used for calculating the equilibrium, stability, and dynamics of fusion plasmas, but has also been used for a variety of other applications, including astrophysical applications. In particular, \textit{M3D-C1} is designed to address some of the most critical challenges confronting tokamak plasmas: large-scale instabilities, which significantly degrade thermal confinement; and disruptions, which cause complete loss of energy confinement and which may cause damage to reactor-scale tokamaks.
\textit{M3D-C1} builds upon some of the design principles pioneered by the \textit{M3D} code, but the two codes are developed independently and do not share source code. The ``C1" in \textit{M3D-C1} refers to the $C^1$ continuity property of its finite elements, which ensures that both the value and the derivatives of fields are continuous across mesh element boundaries.
Advanced numerical methods are employed in \textit{M3D-C1} to permit the efficient solution of its model equations over a broad range of temporal and spatial scales. These methods include the use of high-order finite elements on an unstructured mesh; fully implicit and semi-implicit time integration options; and parallelization via domain decomposition and the use of scalable parallel sparse linear algebra solvers.
\textit{M3D-C1} is used by over a dozen physics groups around the world including university, national laboratory, and industry users.


\textit{Impact:} The advanced time integration scheme in \textit{M3D-C1} results in about 10 linear system solves per time step depending on the model. The base equations are given in Appendix \ref{sec:appendix} and further details can be found in \cite{ferraro2008nonideal}.
These solves lie on a broad spectrum of difficulty, with the momentum solve being the most challenging and least robust.
This work focuses on the momentum solve and only on the PETSc solver component (e.g., not matrix assembly).
The solver developed here is, however, fully applicable to ``hard" solves in \textit{M3D-C1}.
Many of the solves are mass dominated and easy to solve.
The momentum solve accounts for about $3\%$ of the total run time, but is the most expensive, least robust and the starting point for developing fast solvers in \textit{M3D-C1}.

\textit{Multigrid} (MG) methods solve a fine-grid problem by leveraging a hierarchy of coarse resolution grids. On each grid, coarse function spaces are constructed, and residuals and coarse grid corrections are transferred between levels via restriction and interpolation operators. 
At each level, inexpensive solvers, known as \textit{smoothers}, are used to reduce high-frequency error components, on that grid, and the coarse grid solver must eliminate all error.
MG can provably solve the Laplacian, \textit{the model problem}, to discretization error accuracy with a work complexity of about five residual calculations, using the full MG cycle \cite{UTrottenberg_CWOosterlee_ASchueller_2000a}.
Achieving this so-called \textit{textbook multigrid efficiency} (TME) on a given problem requires understanding and exploiting the structure of the equations and discretizations.
A variety of multigrid methods have been developed \cite{UTrottenberg_CWOosterlee_ASchueller_2000a} and TME has been achieved to some extent on MHD problems \cite{Adams2010} and CFD problems \cite{thomasdiskinbrandt2001}.

\textit{The dynamics of tokamak fusion plasmas are highly anisotropic}, dominated by magnetic guide fields that confine the charged particles to trajectories around the torus (Figure \ref{fig:stell}).
All tokamak models exploit this structure with a \textit{poloidal plane} grid that is approximately perpendicular to the magnetic guide field.
These $3D$ domains are invariably tensor products of $2D$ poloidal plane grids (and discretizations) with a $1D$ toroidal grid.
Many tokamak codes use unstructured grids in the poloidal plane that are thus ``extruded" around the torus.
The structured grid in the toroidal direction suggests the use of a standard approach to anisotropy on structured grids \cite{UTrottenberg_CWOosterlee_ASchueller_2000a}, geometric semi-coarsening, which is the focus of this paper.
Applying MG to the poloidal plane is the subject of future work.

\textit{The $3D$ velocity field in M3D-C1 is decomposed} into three scalar functions corresponding to three waves of the MHD system (Appendix \ref{sec:momentum}).  
This model can include one, two, or all three waves.
Using all three is common in practice, but reduced systems are useful for verification. With this decomposition, on a Cartesian grid without nonlinear terms, the three scalar equations decouple. 
In practice the variables are weakly coupled, see Appendix \S\ref{sec:blocknorms} for quantitative examples.
Because of this weak coupling, \textit{FieldSplit} preconditioners in \textit{PETSc} that treat each variable separately in a (3) block Gauss-Seidel iteration are worth investigating.
\textit{FieldSplit} preconditioners support custom blocks for each wave, which could be effective given that these waves have different anisotropies.
This is a subject of future work.

\textit{These multigrid ideas are fairly generic} and some of them are applicable to other production MHD tokamak codes such as \textit{NIMROD}\footnote{\url{https://nimrodteam.org}}, which uses a different formulation of the extended MHD system that also exploits the toroidal structure of tokamaks, but with a spectral decomposition in the toroidal coordinate.
The resulting matrix is also a block diagonal matrix under ideal, linear conditions, but the number of blocks is much larger than three, being the number of toroidal modes kept in the system, on the order of $30-50$.
Thus \textit{NIMROD} exploits the structure that we exploit with $1D$ multigrid at the formulation level.
The resulting $2D$ linear system is similar to some extent to the poloidal plane solve in \textit{M3D-C1} in that the matrix is likely nearly block diagonal and both use high-order accurate finite element discretizations of unstructured $2D$ poloidal grids.
This structural similarity suggests that the two future work topics in this paper, geometric MG in the poloidal plane and custom small block Jacobi smoothers, could be broadly applicable to \textit{NIMROD}.

\textit{The main contribution }of this paper is demonstrating that multigrid methods can provide a path toward faster, scalable, and robust solvers for even the most complex and ill-conditioned of science and engineering relevant magnetized plasma models and that semi-coarsening in toroidal coordinates is a useful first step in this process for common tokamak codes.
This work proceeds by describing the \textit{M3D-C1} velocity evolution equations and solver methods in \S\ref{sec:eqs}, the model problems for performance evaluation and verification are defined in \S\ref{sec:examples}, performance results are presented in \S\ref{sec:perf} and \S\ref{sec:future} concludes.

\section{\label{sec:eqs} \textit{M3D-C1} equations, multigrid and solver structure}

\textit{Multigrid design methodology:} While iterative solvers have the potential to be effective PDE solvers, they require an understanding of the equations -- and their discretizations -- to achieve this potential \cite{Adams2010,thomasdiskinbrandt2001,Chacon2008}.
Given the equations and discretizations, analysis techniques are available to guide the design of, and even prove the optimality of, multigrid methods.
Examples of such tools are local Fourier analysis (p. 98 \cite{UTrottenberg_CWOosterlee_ASchueller_2000a}), $h$-ellipticity (p. 121 \cite{UTrottenberg_CWOosterlee_ASchueller_2000a}), principal component analysis, which in this context refers to writing the equations in matrix form, taking the determinant, and ordering the terms to understand the dominant terms of the system (e.g., whether it is elliptic).

\subsection{M3D-C1 equations}
Expressing the explicit algebraic form of the \textit{M3D-C1} model for analysis is not feasible and the methods used here are generic for elliptic operators.
\textit{M3D-C1} uses $2D$ unstructured grids in the poloidal plane that are extruded in the toroidal direction.
These prism elements use a $2D$ $C^1$-continuous Bell element in the poloidal plane \cite{Bell69}, together with cubic Hermite elements in the toroidal direction \cite{Jardin04}.
These discretizations include not only the variable itself as a degree of freedom (DoF), the zeroth derivative, but also higher derivatives for a total of 12 equations per physical DoF.

\textit{Parabolization} is a transformation of the extended MHD equations in \textit{M3D-C1} (Appendix \ref{sec:parb}) \cite{HARNED198657,JARDIN2012822,Goedbloed_Poedts_2004,Freidberg1987}.
The momentum equation is parabolized by taking a time derivative and then eliminating the resulting time derivatives of density and magnetic field using the ideal MHD equations. 
This results in two fourth order, symmetric, curl operators ($U, \chi$) and one Laplacian ($\omega$), discretized with high order, $C^1$-continuous finite elements.
Symmetry and ellipticity are not necessary to successfully apply multigrid to resistive MHD \cite{Adams2010}, but they are very useful properties and critical for solving these problems with multigrid.

\subsection{M3D-C1 solvers}
\label{sec:solvers}

\textit{M3D-C1 solvers} use \textit{PETSc}\footnote{\url{https://petsc.org}} Krylov solvers (FGMRES or GMRES) and employ a block Jacobi (BJ) preconditioner, with exact subdomain solves, in which each block corresponds to a poloidal plane.
This is a natural decomposition of the problem given that an unstructured grid is used in the poloidal plane to capture the geometry of the tokamak wall and magnetic flux surfaces.
This $2D$ grid is extruded with a regular $1D$ grid in the toroidal direction.
The structured toroidal grid makes semi-coarsening geometric MG particularly simple, with geometric 2:1 coarsening and a 3-point stencil\footnote{$\left] \frac{1}{2} \quad 1 \quad \frac{1}{2} \right[$} prolongation operator $P$.
Having the application compute Jacobians on coarse grids is not practical, but given $P$ the coarse grid operators can be constructed algebraically with a Galerkin process: $A_{i+1} = P^TA_i P$.
This results in a geometric/algebraic multigrid solver with a purely algebraic interface \cite{Adams-99a}.

\textit{The PETSc solver interface in M3D-C1} takes matrix and vector objects, populated by the application in an object-oriented interface.
The numerics are performed with high level controls on the host and distributed linear algebra back-ends, some of which are third-party libraries like parallel direct solvers \textit{MUMPS} and \textit{SuperLU}, and device linear algebra libraries like \textit{cuSPARSE}, as well as non-numerical methods such as mesh partitioners for load balancing.
Stand-alone PETSc has built-in host linear algebra and geometric and algebraic MG preconditioners.
All data can remain device resident with time integrators, nonlinear solvers and Krylov linear solvers operating through abstract interfaces to the linear algebra primitives that run on the device.
A solver object is then created, the solver is called, and the solution is extracted from the PETSc solution vector.
Solver parameters can be set in the code, but using command-line parameters is recommended for runtime flexibility (see Appendix \ref{sec:params} for an example).

\textit{M3D-C1} computes one new Jacobian matrix every time step, without a nonlinear iteration.
Furthermore, the preconditioner is lagged and with a \textit{refresh period} provided by the user because the setup costs of matrix factorizations, etc., are significant.
This \textit{setup}, or refresh, for the production BJ solver is primarily full LU factorizations on each plane.
The MG solver also requires a full plane LU factorization for the coarse grid and all grids when the BJ solver is used as the smoother.
Additionally, MG has a cost in the Galerkin coarse grid construction.
As the physics evolves, solver convergence degrades from the lack of consistency between the solver and matrix.
This refresh model is very simple and a smarter algorithm could clearly be useful in these problems where the physics appears to be reaching a quasi steady state and the refresh period could be increased dramatically.

\textit{Solver tolerance.} The relative solver tolerance, $rtol=\frac{|b - A \tilde{u}|}{|b|}$, is the primary parameter that governs solver accuracy and cost.
The \textit{M3D-C1} team has determined that an $rtol=10^{-9}$ is reliable and is used in production.
For the purpose of experiments, we wish to optimize the solver by reducing the tolerance by studying accuracy in a way that would not be practical for users.
\textit{M3D-C1} reports nine quantities (kinetic and magnetic energies, etc.) at each of the 100 time steps in the test problem.
The verification test compares this data to data collected with a high-accuracy solver tolerance.
The test is deemed to pass if each value, $x$, and its counterpart in the high-accuracy data, $\hat x$, has a {relative} difference $\frac{|x - \hat x|}{|\hat x|}$ less than $10^{-3}$.
We modified this test to filter out false positives from $\hat x$ values that are near zero by first scanning for the largest absolute value for each of the nine quantities for use as $\hat x$.
The failure tolerance is reduced to $10^{-2}$ and the number of tests that violated the $10^{-3}$ tolerance is reported.
This test is used to set custom tolerances for the SPARC problem in \S\ref{sec:newverifyresults}.
Data with the production tolerance of $rtol=10^{-9}$ is reported in \S\ref{sec:gpu09} and \S\ref{subsec:stell}.

\subsection{Multigrid background}

Grids are traversed according to various cycling strategies in multigrid, the most common being the \textit{V-cycle}, in which the coarsest-grid solve forms the base of the ``V". Smoothing steps descending the hierarchy (on the left leg of the ``V") are called \textit{pre-smoothing}, while those applied on the ascent (the right leg) are called \textit{post-smoothing}.
A common shorthand for the cycle type and number of pre-smoothing and post-smoothing steps is $V(pre,post)$, e.g., $V(1,1)$ uses a single pre-smoothing and post-smoothing step.
Two smoothers are used in this work: point-block Jacobi (PBJ) smoothers 
and the BJ solver can be repurposed as a smoother.
With $C^1$ elements, all degrees of freedom are on the vertices of the mesh.
The Bell elements are about fifth-order accurate (in plane) and the cubic splines are fourth-order accurate \cite{Jardin04}.
This discretization has 12 DoFs per vertex, per physical zeroth derivative DoF, and with three DoFs in the momentum solve the matrix has a \textit{block size} of 36, that is 36 equations per vertex.
PBJ smoothers in PETSc compute explicit, dense inverses of these diagonal blocks, which can be applied effectively on GPUs with dense matrix-vector products.
A BJ smoother is a powerful and expensive smoother and is used in a $V(0,1)$-cycle for the most challenging problems investigated here.
The PBJ smoother is fast and uses less memory and is useful on some of the problems.

\textit{Previous work on MHD solvers.} While Adler et al.\ have developed multigrid methods for a two field viscoresistive MHD model with curl operators \cite{Adler2020,Adler2019,Adler2016}, and Chac\'{o}n used a parabolization (or Schur complement) approach to apply multigrid to Laplacians in an MHD solver  \cite{Chacon2008}, there is no prior work on applying multigrid to solve systems approaching the complexity of those in \textit{M3D-C1}.

\section{\label{sec:examples} Model test problems}

\begin{figure}[h!]
    \centering
    \includegraphics[width=0.5\linewidth]{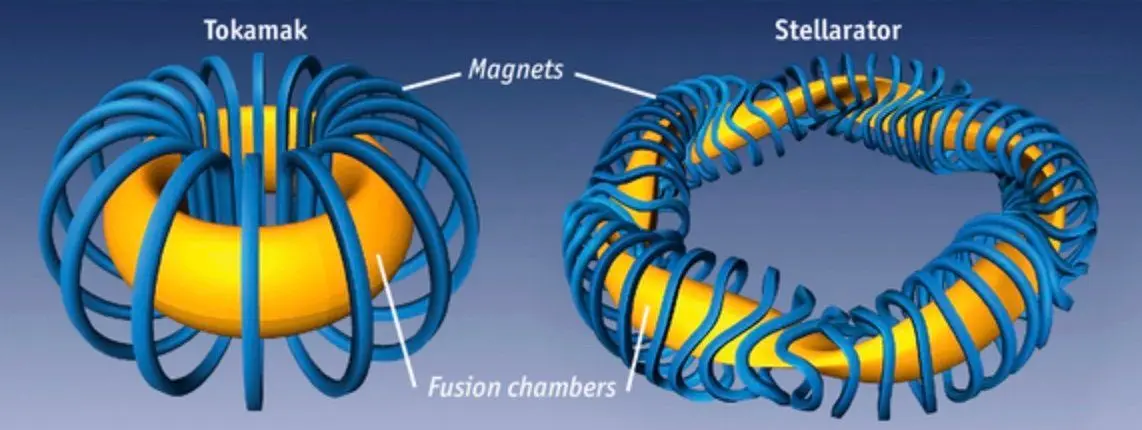}
    \caption{Traditional tokamak and stellarator geometry}
    \label{fig:stell}
\end{figure}

This section describes test problems used for performance evaluation.
Two tests are used: a runaway electron model of a SPARC tokamak disruption, and a stellarator problem.
The stellarator test uses a coordinate transformation in \textit{M3D-C1} for the geometry of a stellarator.
Figure \ref{fig:stell} shows the geometry of a generic tokamak (left) and a generic stellarator (right).


\subsection{SPARC disruption with runaway electrons problem}


SPARC is a compact, high-field ($B_0 = 12.2$\,T),
high-current ($I_p = 8.7$\,MA) tokamak designed to achieve $Q > 2$ in
deuterium--tritium plasmas. \textit{M3D-C1} incorporates a fluid
runaway electron model coupled with the bulk MHD
equations 
~\cite{Datta2025,datta2025modelingeffectmhdactivity}.

\subsection{Quasi-axisymmetric stellarator test}
\label{sec:stell}


The second test is a stellarator problem using M3D-C1's recently developed
coordinate transformation to adapt to non-axisymmetric stellarator
geometry~\cite{zhou2021}.  The configuration is a quasi-axisymmetric (QA)
stellarator optimized for low quasisymmetry error, good energetic particle
confinement, and self-consistent bootstrap current~\cite{saxena2025bootstrapcurrentmodelingm3dc1,landreman2022}, with a
minor radius of 1.70\,m and volume-averaged $B = 5.86$\,T.
Bootstrap current, which represents a significant fraction of the total current density in QA stellarators, is included self-consistently via the Sauter--Redl--Landreman formulation~\cite{sauter1999,redl2021,landreman2022,saxena2025}, which exploits the isomorphism between axisymmetric and quasi-symmetric geometries.
This test case is significantly more challenging for the solver than the SPARC problem: the BJ solver fails to converge (Figure \ref{fig:stell_09}, whereas the multigrid solver
remains effective, albeit using BJ as the smoother, demonstrating the robustness advantage of the MG approach on
non-axisymmetric configurations.
This project is under active development and multiple physicists have observed the BJ solver fail and used the MG solver to continue work.


\section{\label{sec:perf} Performance evaluation}

This section evaluates the performance of the solvers developed in this work on these two fusion modeling problems.
The Perlmutter machine at NERSC is used for all numerical studies.
Perlmutter has CPU nodes with two 64-core AMD EPYC 7763 processors and GPU nodes with one 64-core AMD EPYC 7763 and 4 NVIDIA A100 GPUs.
The 2D poloidal plane mesh is partitioned into processor subdomains in an offline setup phase.
This mesh is extruded around the tokamak or stellarator with a given number of planes.
The MUMPS LU solver is used for all direct solves.

\subsection{SPARC problem with custom solver tolerances}
\label{sec:newverifyresults} 

The SPARC test problem is run with $4-64$ planes in a scaled speedup study using the custom solver tolerance of $rtol = 10^{-6}$ and $10^{-7}$.
Each plane contains 14,200 vertices with 36 DoFs per vertex, which results in 511,200 equations per plane.
The 2D unstructured mesh of the poloidal plane is partitioned into 32 processor subdomains, resulting in 15,975 DoFs per process and about 12,000,000 non-zeros per process.
All tests use 16 processes per CPU socket, which drive 4 GPUs in the GPU tests.
For the CPU study, one plane is placed on each node with two CPU sockets, and for the GPU study two nodes with one CPU socket each are used for each plane, resulting in 8 GPUs per plane and about 64K equations, or 48M non-zeros, per GPU.

The custom tolerance of $rtol=10^{-6}$ in the 4, 8 and 16 plane cases, and $rtol=10^{-7}$ in the 32 and 64 plane cases is used.
There are 11 and 10 warnings with PBJ smoothing and BJ solver, respectively, in the 64 plane case where the $10^{-3}$ tolerance in the verification test (\S\ref{sec:solvers}) was violated.
These warnings occurred in time steps $2-13$, the early growth phase of the test.
Note, this result is interesting in that it indicates that the accuracy at a given solver tolerance is similar, which given the different spectral properties of MG and BJ (MG damps the spectra of the error evenly in theory).
The time step for the 64 plane case is one half that of the other tests, due to CFL constraints.
A refresh period of 20 is used for 4, 8 and 16 plane cases, and a period of 10 is used for the 32 plane case.
In the 64 plane case the MG solver with PBJ smoother was refreshed every time step and the BJ solver and MG solver with BJ smoother were refreshed every 10 time steps.
The $V(1,1)$-cycle PBJ smoother performance was poor in the 64 plane case, which motivated increasing the smoothing by $6x$, which reduces GMRES restarts and improved performance significantly.
The source of this increase in difficulty is not well understood, but it is likely a new nonlinear mode is resolved with increased resolution.

\begin{figure}[h!]
    \centering
    \begin{tabular}{ccc}
        \begin{subfigure}{0.32\textwidth}
            \centering
            \includegraphics[width=\linewidth]{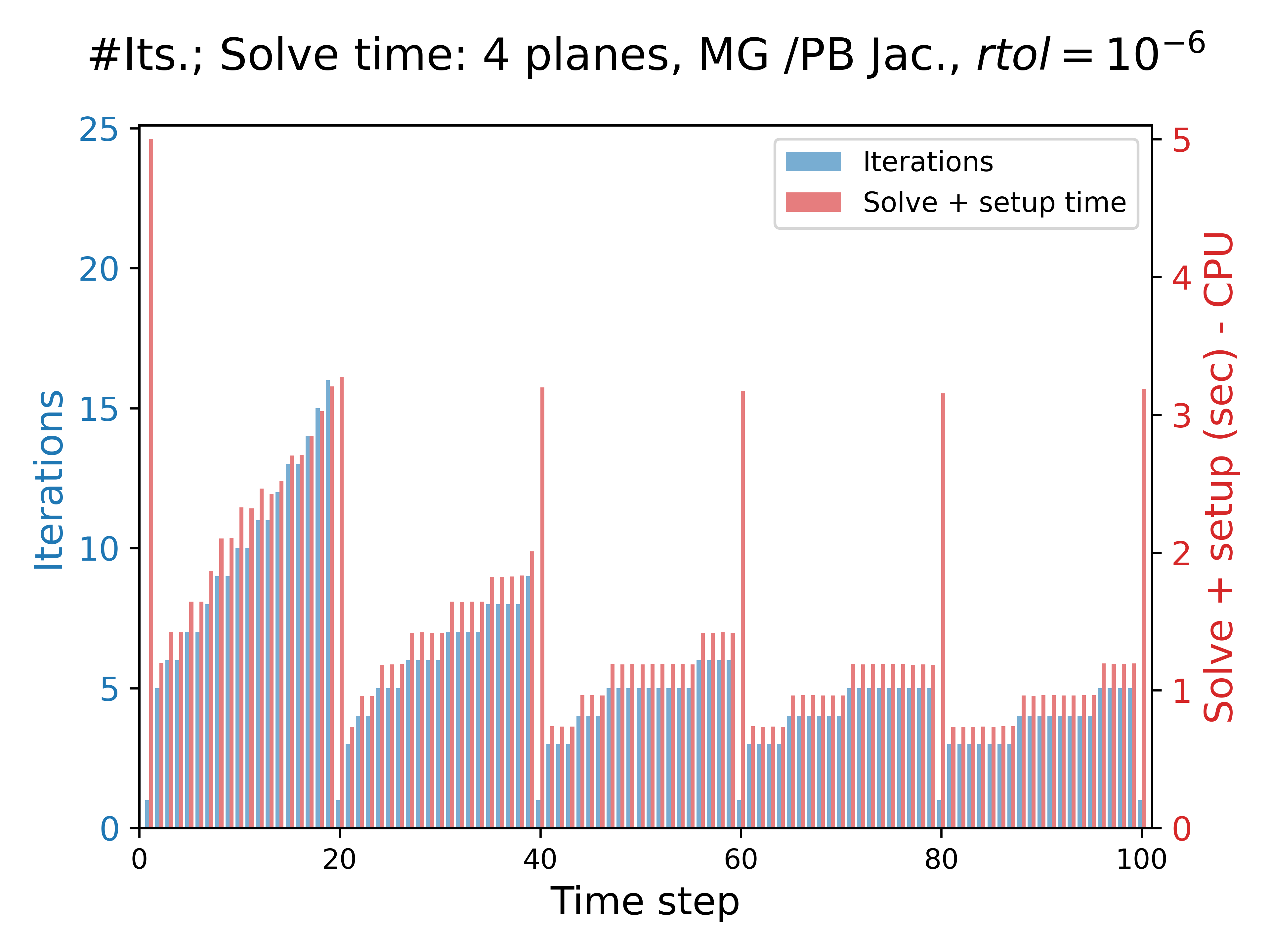}
            \caption{MG $V(1,1)$ PBJ smoother}
            \label{fig:4-MG}
        \end{subfigure} &
        \begin{subfigure}{0.32\textwidth}
            \centering
            \includegraphics[width=\linewidth]{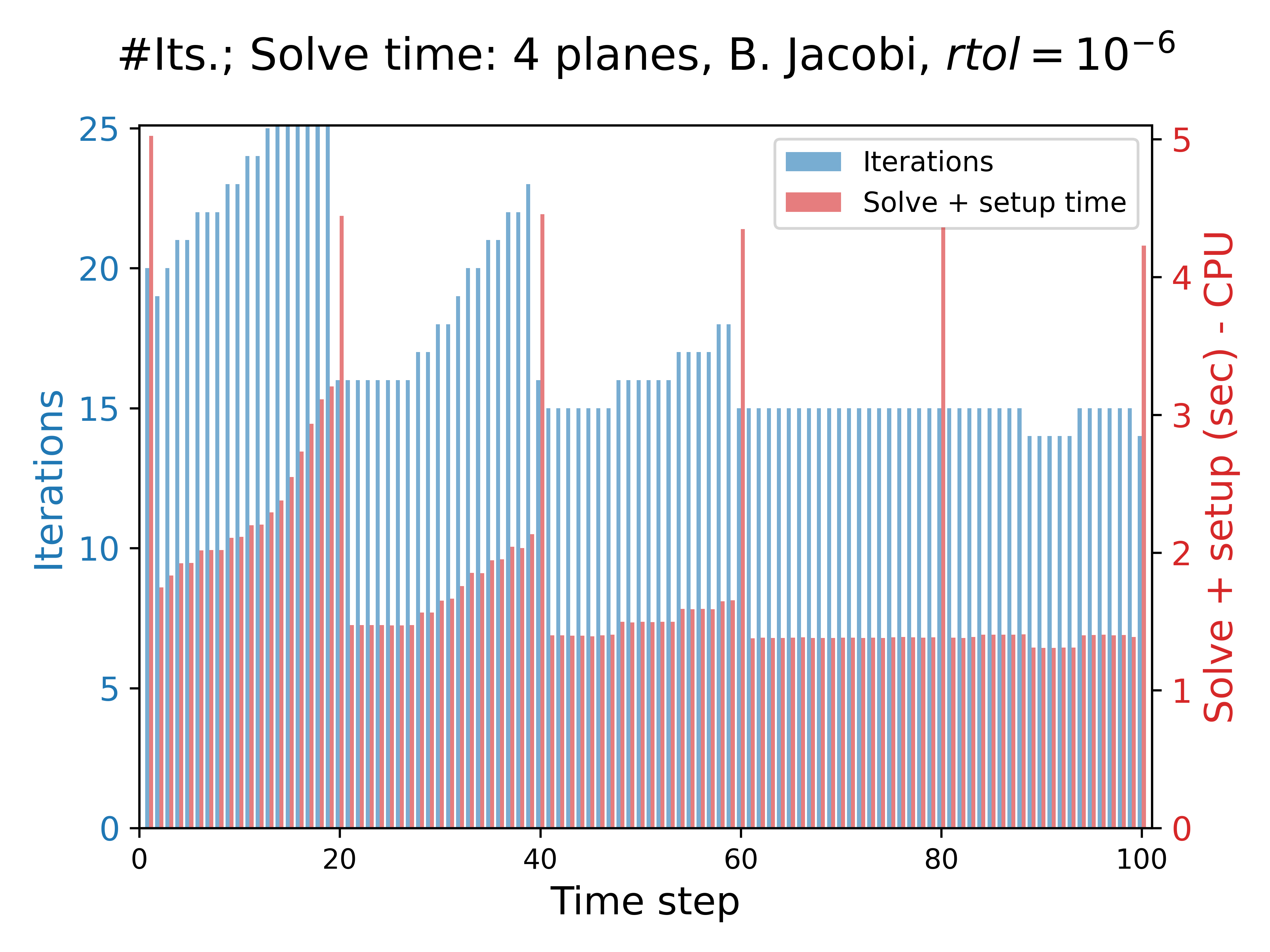}
            \caption{Block Jacobi}
            \label{fig:4-BJ}
        \end{subfigure} &
        \begin{subfigure}{0.32\textwidth}
            \centering
            \includegraphics[width=\linewidth]{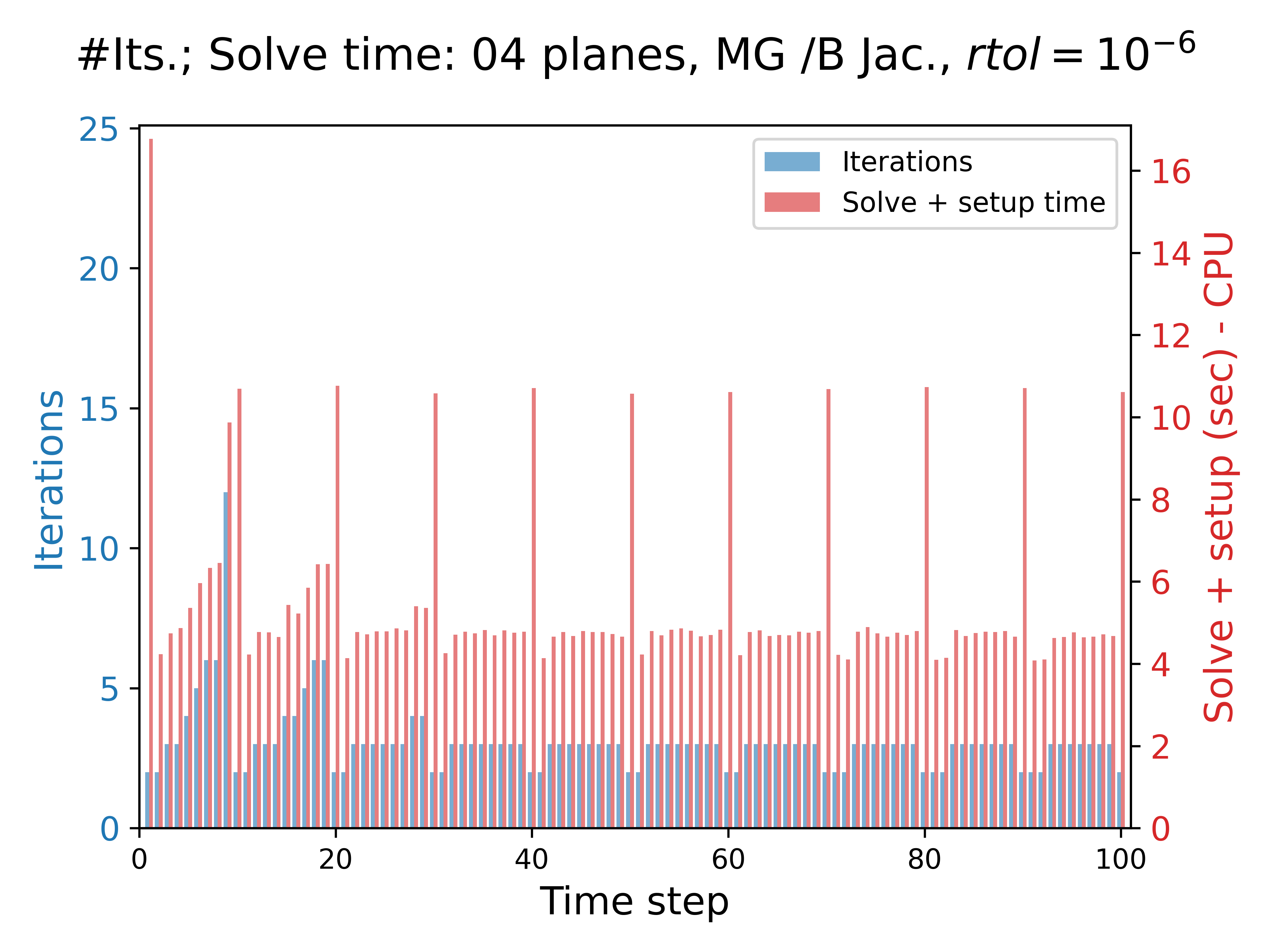}
            \caption{MG $V(0,1)$ Block Jacobi}
            \label{fig:4-mgBJ}
        \end{subfigure} \\[1ex]
    \end{tabular}
  \caption{4 plane case, iteration counts (left axis/bar) and solve times with setup (right axis/bar)}
  \label{fig:4_pl}
\end{figure}
\begin{figure}[h!]
    \centering
    \begin{tabular}{ccc}
        \begin{subfigure}{0.32\textwidth}
            \centering
            \includegraphics[width=\linewidth]{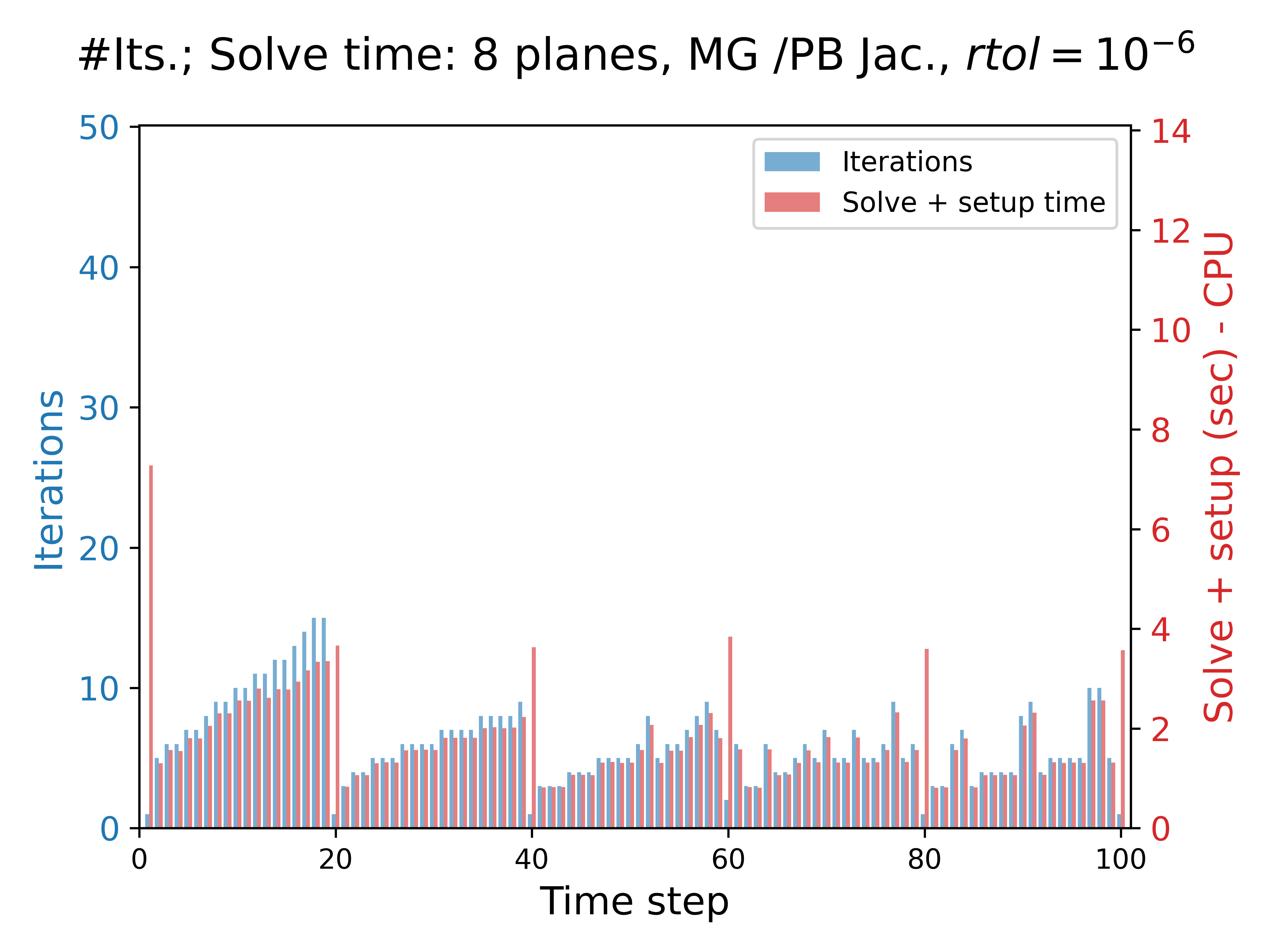}
            \caption{MG $V(1,1)$ PBJ smoother}
        \end{subfigure} &
        \begin{subfigure}{0.32\textwidth}
            \centering
            \includegraphics[width=\linewidth]{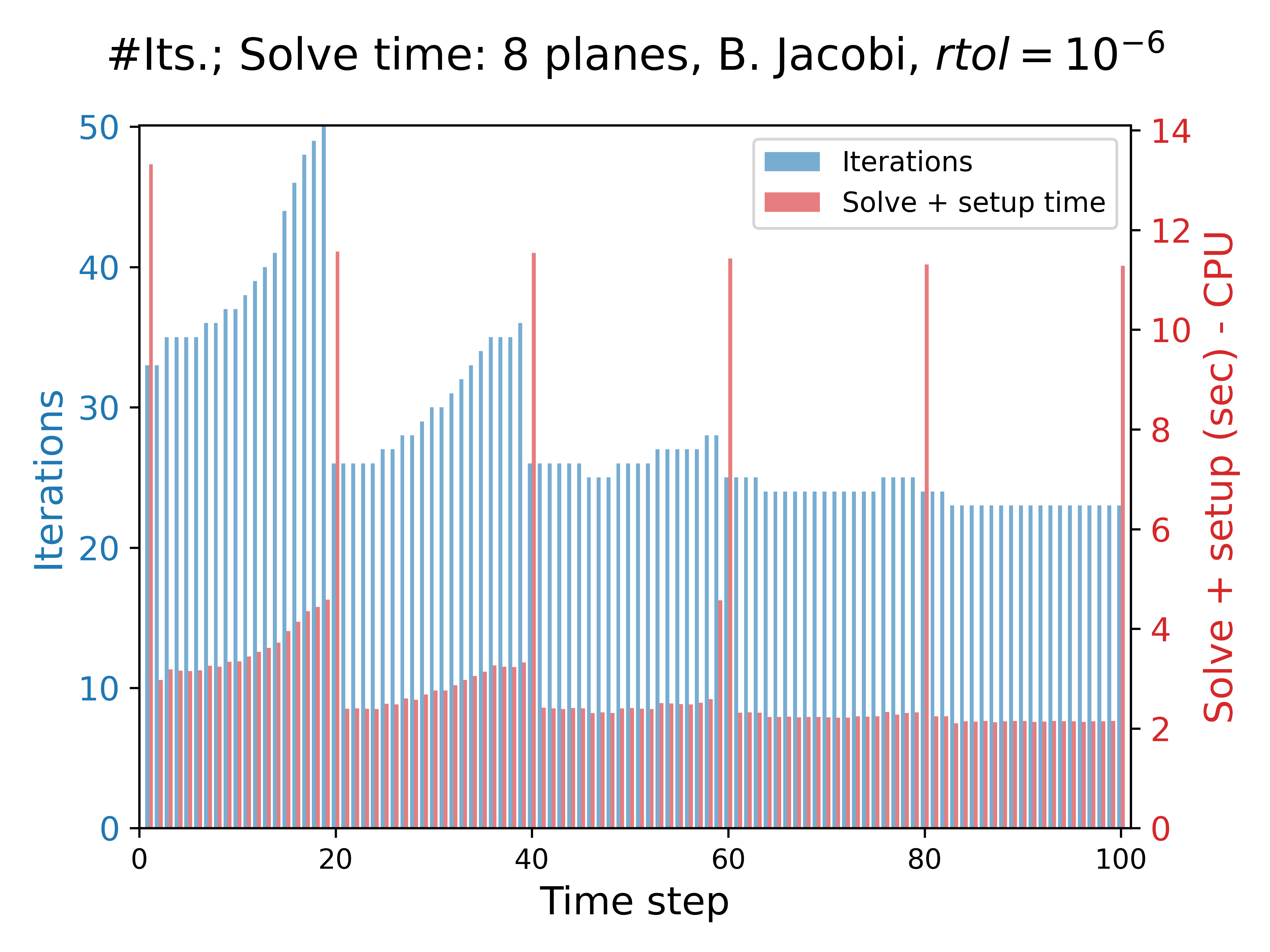}
            \caption{Block Jacobi}
        \end{subfigure} &
        \begin{subfigure}{0.32\textwidth}
            \centering
            \includegraphics[width=\linewidth]{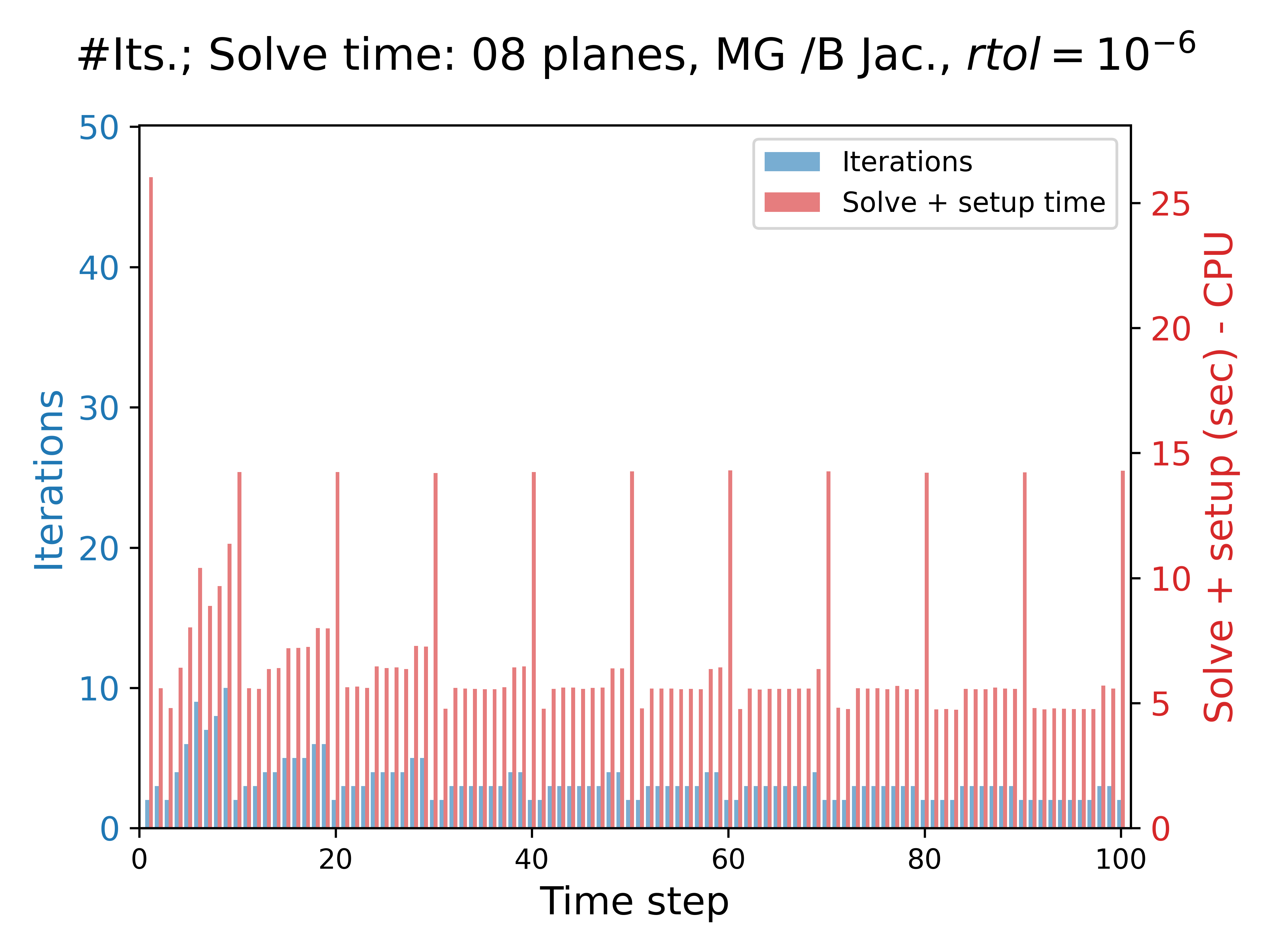}
            \caption{MG $V(0,1)$ Block Jacobi}
        \end{subfigure} \\[1ex]
    \end{tabular}
    \caption{8 plane case, iteration counts (left/blue), solve times with setup (right/red) vs time step}
    \label{fig:8_pl}
\end{figure}
\begin{figure}[h!]
    \centering
    \begin{tabular}{ccc}
        \begin{subfigure}{0.32\textwidth}
            \centering
            \includegraphics[width=\linewidth]{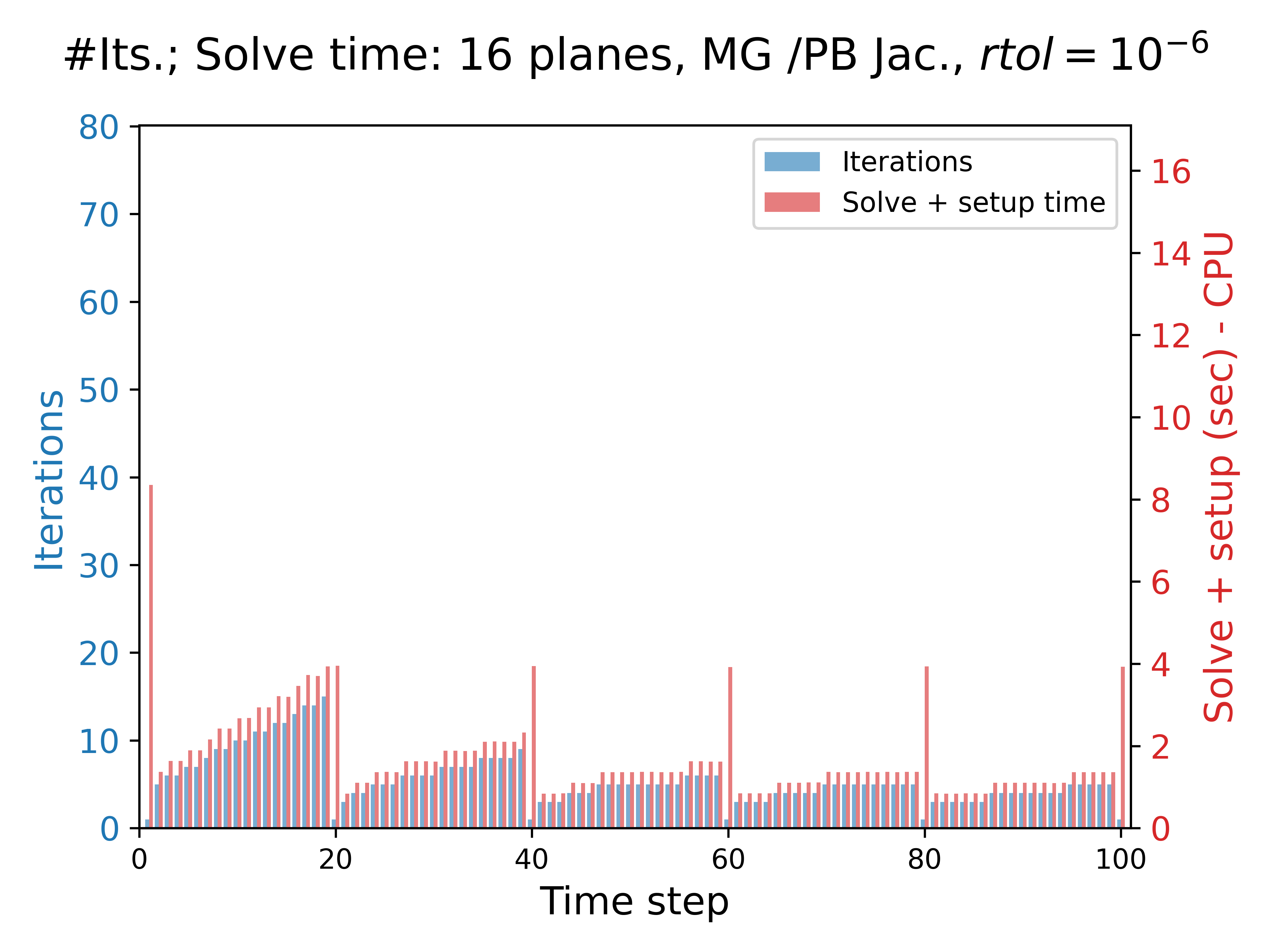}
            \caption{MG $V(1,1)$ PBJ smoother}
        \end{subfigure} &
        \begin{subfigure}{0.32\textwidth}
            \centering
            \includegraphics[width=\linewidth]{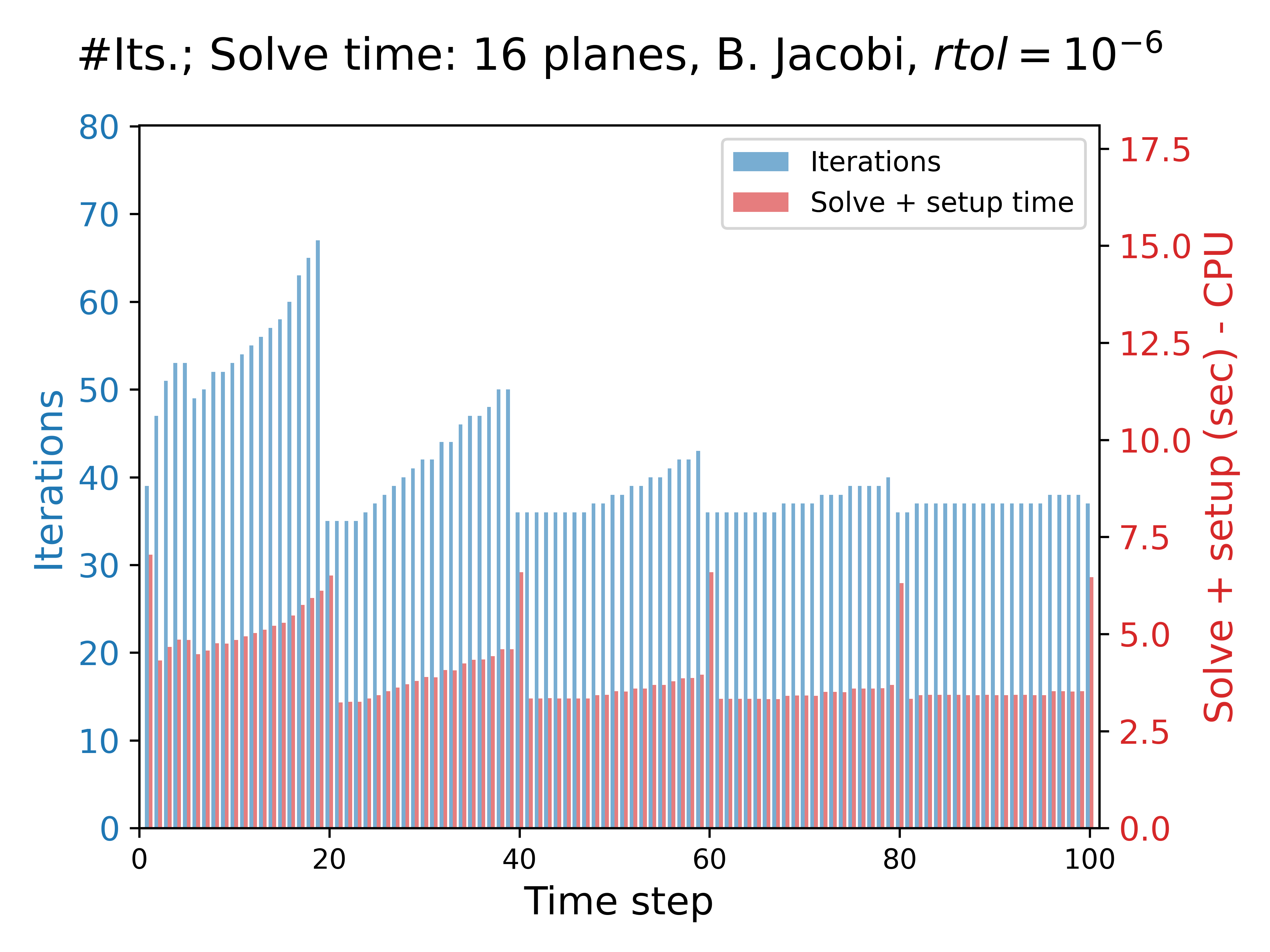}
            \caption{Block Jacobi}
        \end{subfigure}  &
        \begin{subfigure}{0.32\textwidth}
            \centering
            \includegraphics[width=\linewidth]{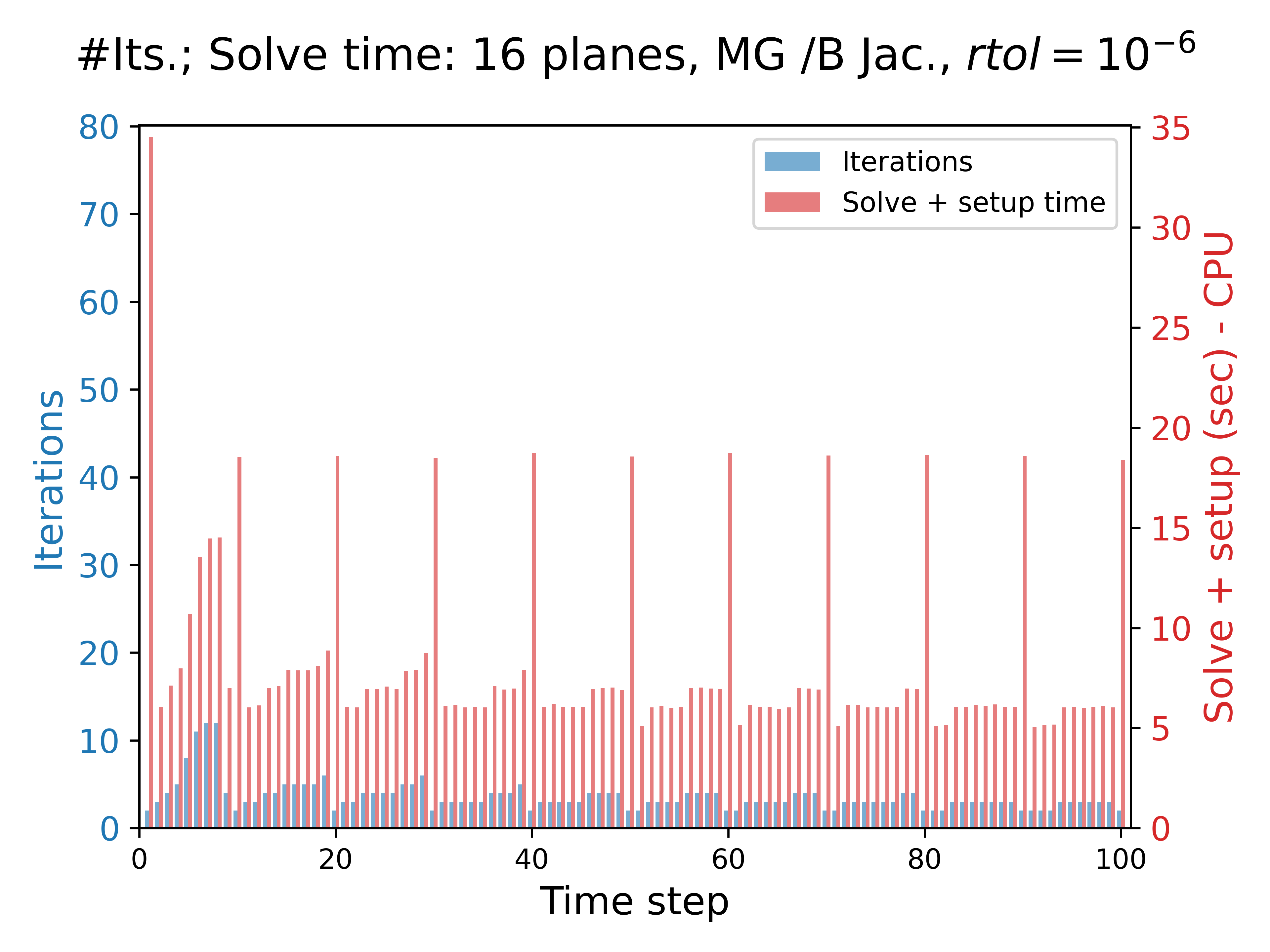}
            \caption{MG $V(0,1)$ Block Jacobi}
        \end{subfigure} \\[1ex]
    \end{tabular}
  \caption{16 planes case, iteration counts (left/blue), solve times with setup (right/red) vs time step}
  \label{fig:16_pl}
\end{figure}
\begin{figure}[h!]
    \centering
    \begin{tabular}{ccc}
        \begin{subfigure}{0.32\textwidth}
            \centering
            \includegraphics[width=\linewidth]{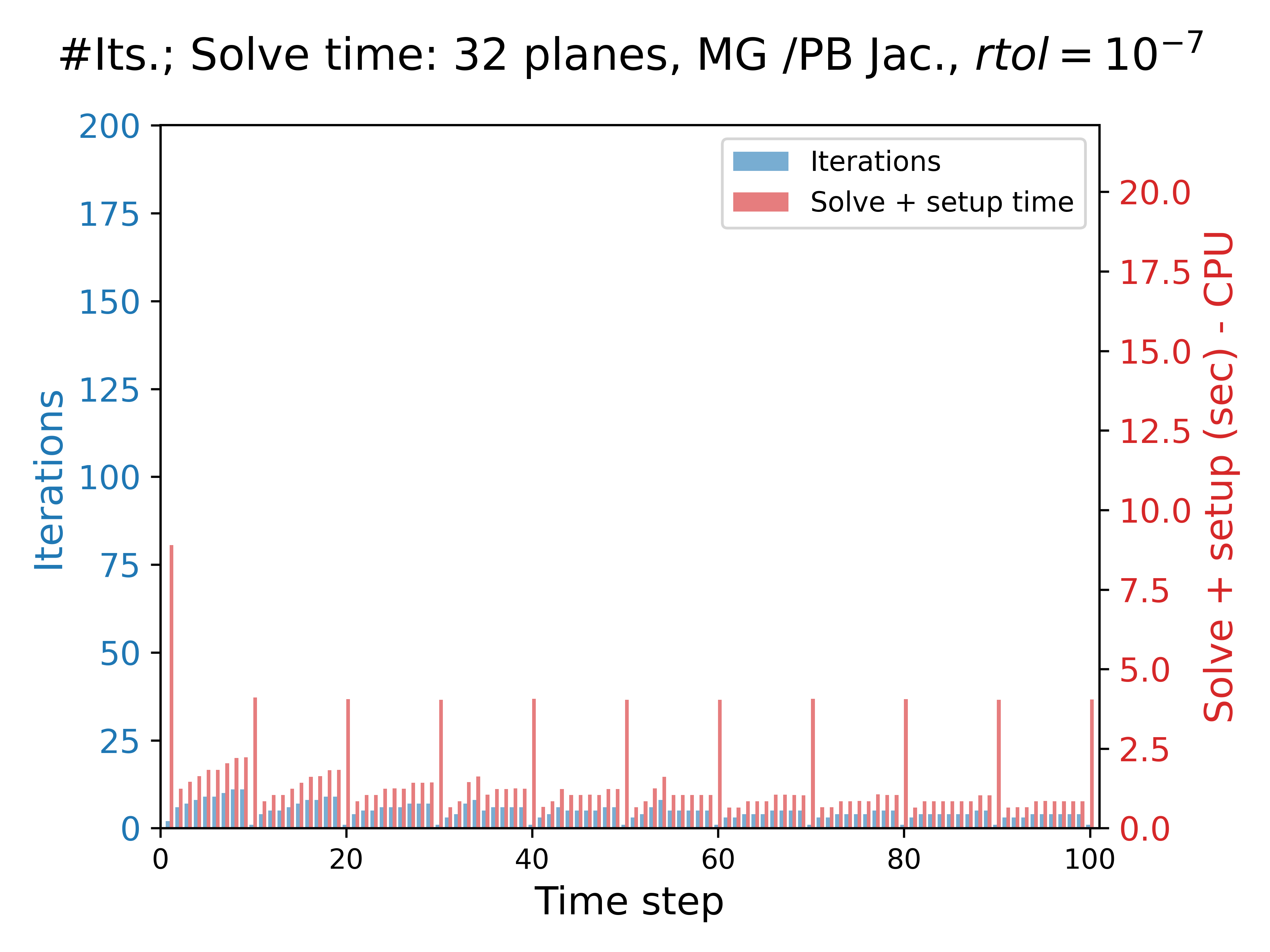}
            \caption{MG $V(1,1)$ PBJ smoother}
        \end{subfigure} &
        \begin{subfigure}{0.32\textwidth}
            \centering
            \includegraphics[width=\linewidth]{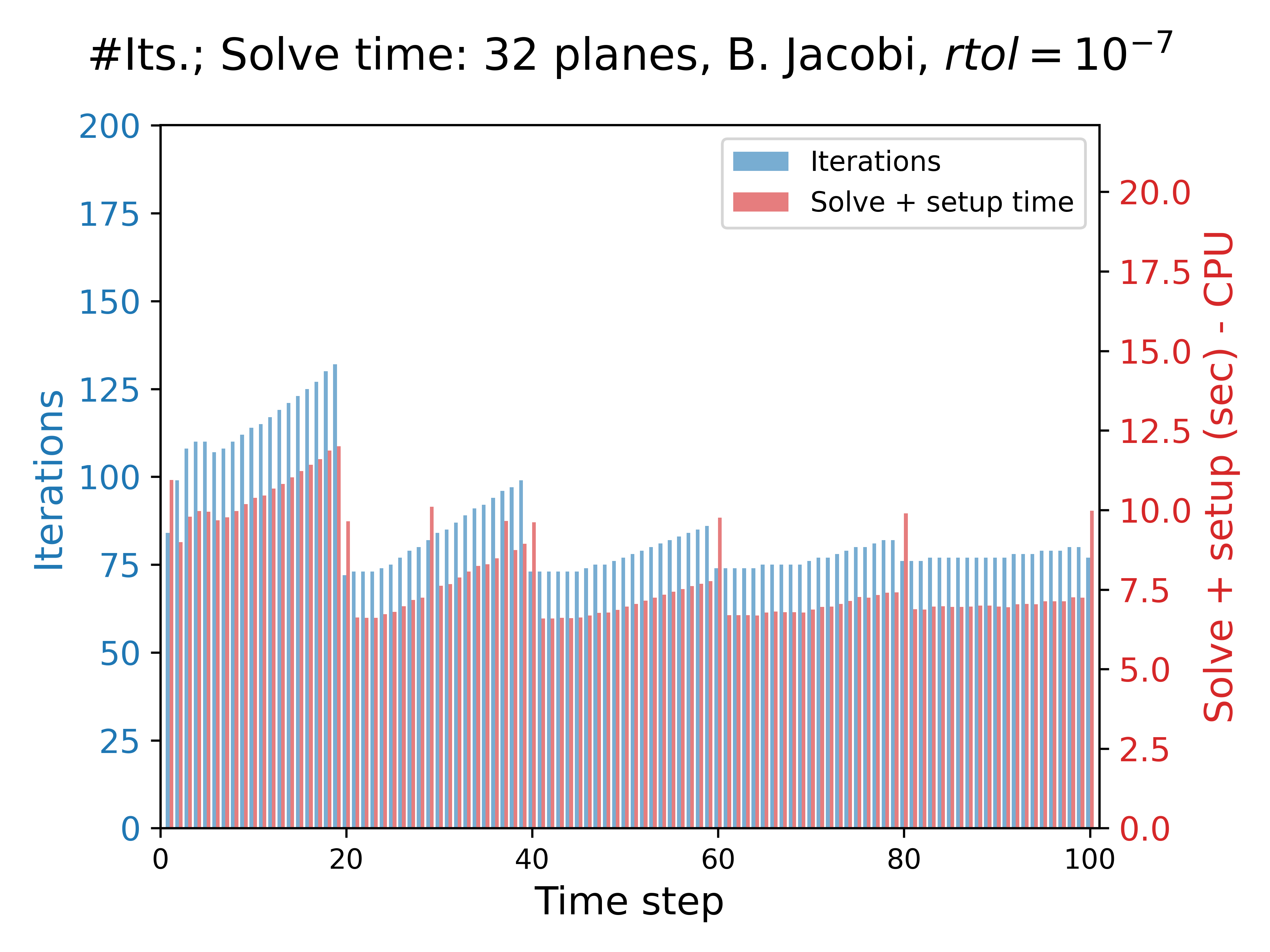}
            \caption{Block Jacobi}
        \end{subfigure} &
        \begin{subfigure}{0.32\textwidth}
            \centering
            \includegraphics[width=\linewidth]{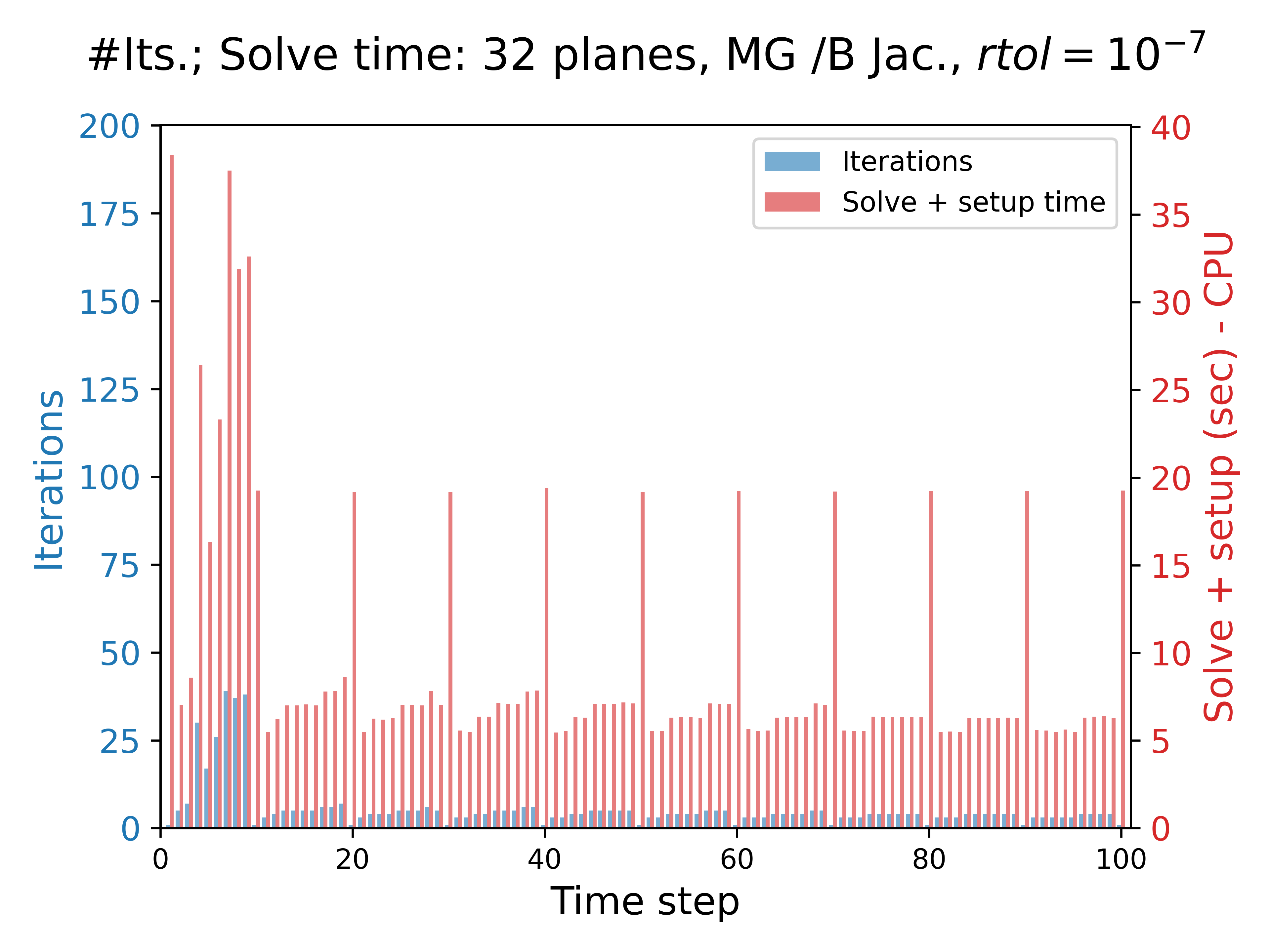}
            \caption{MG $V(0,1)$ Block Jacobi}
            \label{fig:32_mgbj}
        \end{subfigure} \\[1ex]
    \end{tabular}
  \caption{32 plane case, iteration counts (left/blue), solve times with setup (right/red) vs time step}
  \label{fig:32_pl}
\end{figure}
\begin{figure}[h!]
    \centering
    \begin{tabular}{ccc}
        \begin{subfigure}{0.32\textwidth}
            \centering
            \includegraphics[width=\linewidth]{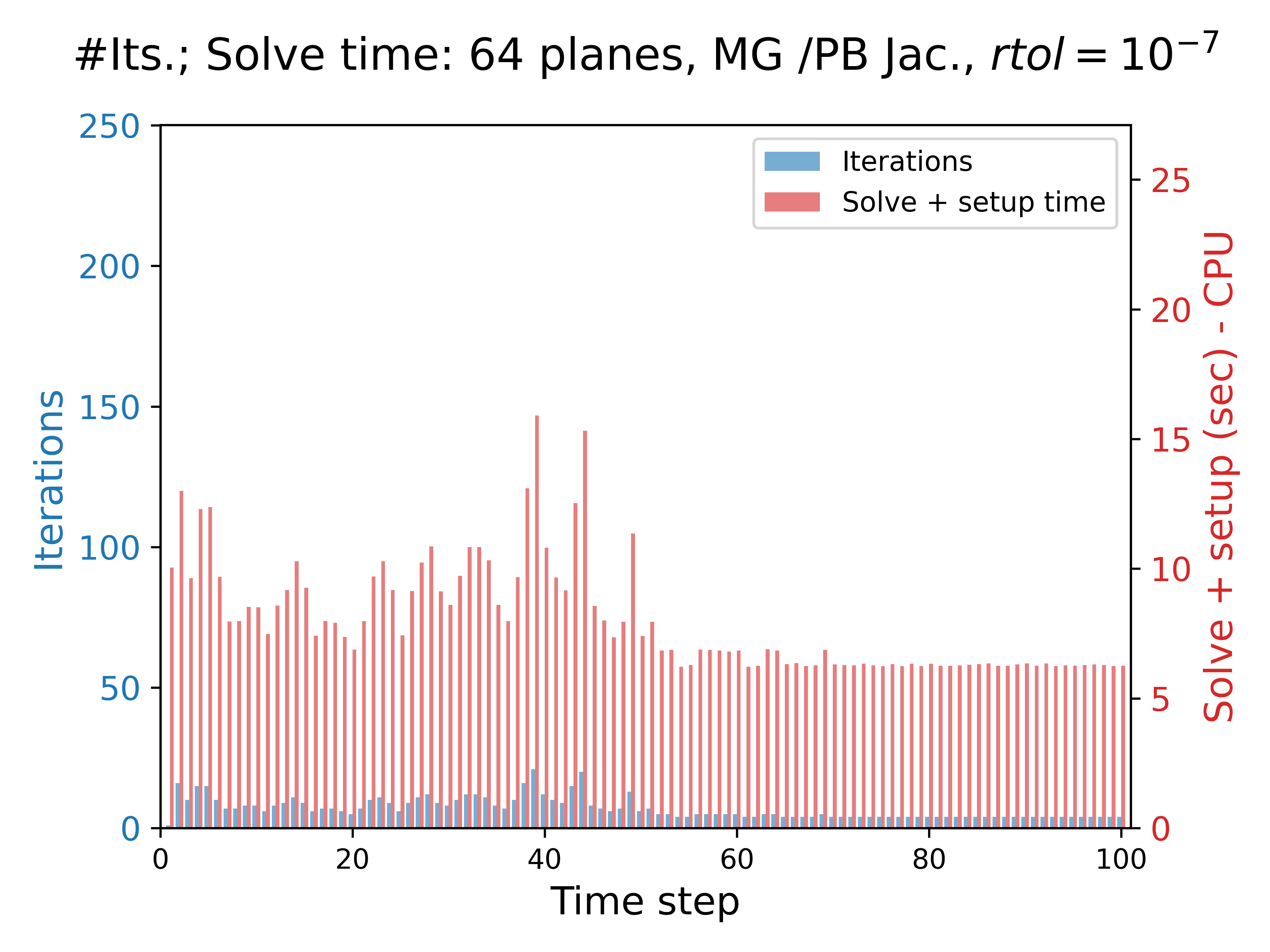}
            \caption{MG $V(6,6)$-cycle PBJ refresh=1}
            \label{fig:64_pl_mg}
        \end{subfigure} & 
        \begin{subfigure}{0.32\textwidth}
            \centering
            \includegraphics[width=\linewidth]{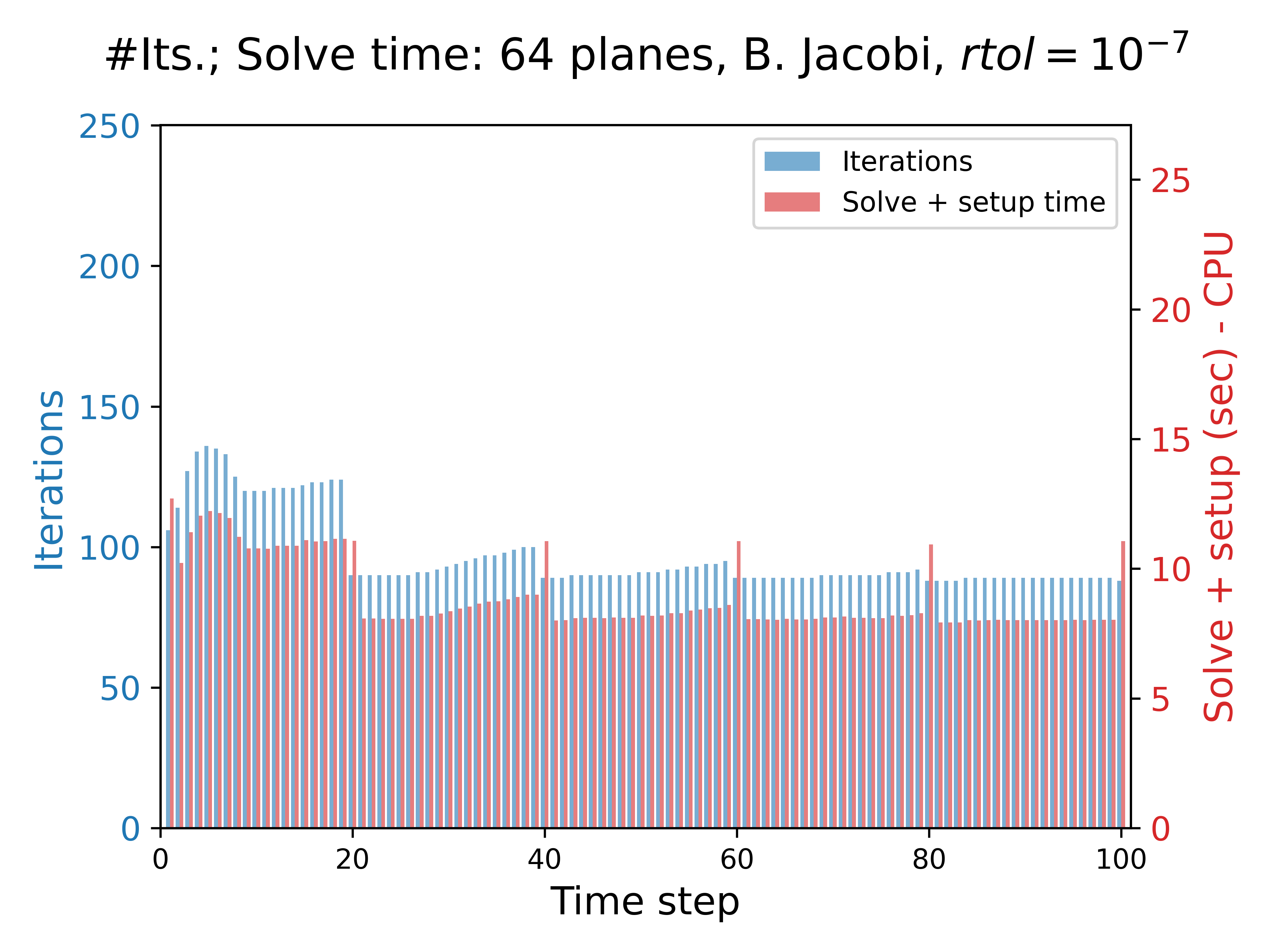}
            \caption{Block Jacobi}
        \end{subfigure} & 
        \begin{subfigure}{0.32\textwidth}
            \centering
            \includegraphics[width=\linewidth]{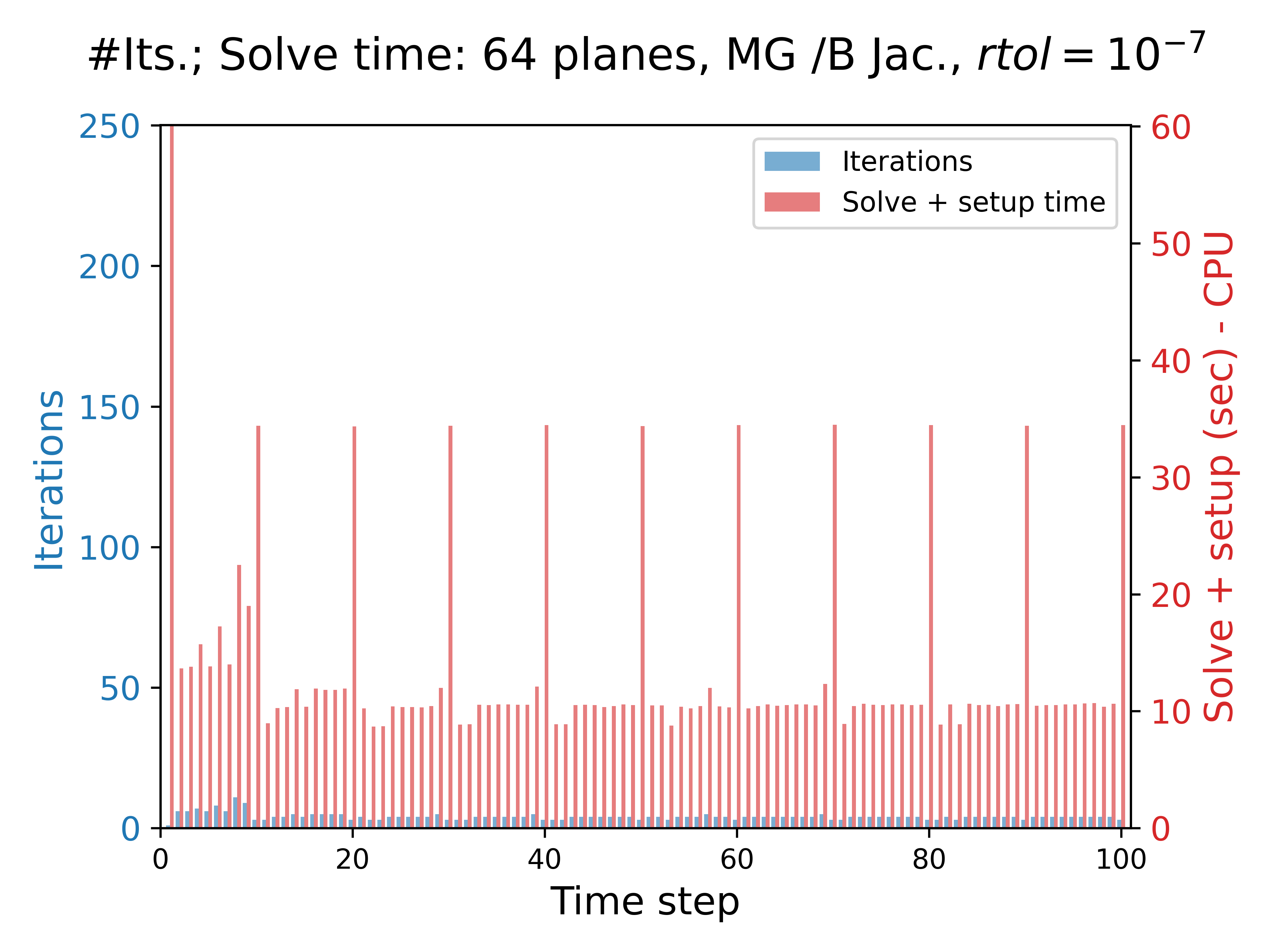}
            \caption{MG $V(0,1)$ Block Jacobi}
            \label{fig:64_pl_v01}
        \end{subfigure} \\[1ex]
    \end{tabular}
  \caption{64 plane case, iteration counts (left/blue), solve times with setup (right/red) vs time step}
  \label{fig:64_pl}
\end{figure}

Figures \ref{fig:4_pl}--\ref{fig:64_pl} show the iteration count (left axis) and solve time (right axis) that includes the periodic solver setup, for each time step as reported by the \textit{M3D-C1} timers. 
Each case is run on CPU nodes with the MG $V(1,1)$-cycle solver with PBJ smoothing, with the production block Jacobi solver, and with the MG $V(0,1)$-cycle with the BJ smoother.
Note, one iteration of GMRES preconditioned with PBJ as the pre-smoother is used with BJ post-smoothing and is denoted as a $V(0,1)$-cycle for clarity.



The $V(0,1)$-cycle with BJ smoothing is not competitive (note the different time scales) and is more sensitive to inconsistency with the operator.
In the 32 plane case in Figure \ref{fig:32_mgbj} the $V(0,1)$-cycle takes more iterations than the $V(1,1)$-cycle with PBJ smoothing even though it is a uniformly more powerful (expensive) solver.
This is not well understood but it can be attributed to the inconsistent approximation ``shooting" the solution further in the wrong direction.

Some observations from this data are as follows:
\begin{itemize}
    \item The MG solver converges quickly when the preconditioner is refreshed or consistent with the operator, converging in one iteration in the $4-16$ plane cases. 
    \item The MG convergence degrades more dramatically than block Jacobi's without a refresh;
    \item Block Jacobi solver iteration counts increase with number of poloidal planes, as expected;
    \item The 32 and 64-plane cases are noticeably harder and a fully refreshed MG solve no longer converges in one iteration and a tighter solver tolerance is required to satisfy the custom verification test;
    \item The 64-plane case is much harder than the 32 plane case, motivating the use of a $V(6,6)$-cycle and a solver refresh every time step;
\end{itemize}

\subsubsection{Scaled speedup data}
\label{ssec:weak}
Table \ref{tab:step_times} shows the velocity solve stage time, as measured with \textit{M3D-C1} timers, which is a superset of the PETSc's ``KSPSolve" solve times shown in Table \ref{tab:it_times}.
\begin{table}[h!]
    \centering
    \begin{tabular}{l|ccccc}
    \hline
    Number of planes       & 4 & 8 & 16 & 32 & 64 \\
    \hline
    Multigrid / PBJ  &  147 & 181 & 176 & 188 & 825$^*$ \\
    Block Jacobi   & 181 & 320 & 516 & 1,293 & 1,704 \\
    \hline 
    \end{tabular}
    \caption{Total velocity solve times (sec) in the \textit{M3D-C1} solve phase timers, $^*V(6,6)$-cycle}
    \label{tab:step_times}
\end{table}
The PETSc solve time and total number of solver iterations over the 100 time steps, for both the multigrid / PBJ and block Jacobi solvers, is shown in Table \ref{tab:it_times}.
\begin{table}[h!]
    \centering
    \begin{tabular}{l|ccccc}
    \hline
    \# planes      & 4 & 8 & 16 & 32 & 64\\
    \hline
    Multigrid / PBJ  &  135 [551] & 163 [615] & 156 [551] & 154 [476] & 772 [430]$^*$ \\
    Block Jacobi     & 164 [1,767] & 264 [2,848] & 498 [5,585] & 1,261 [14,055] & 1,671 [18,353] \\
    \hline
    \end{tabular}
    \caption{Total solver time including setup (sec) [total iterations]  PETSc ``KSPSolve" time, $^*V(6,6)$-cycle}
    \label{tab:it_times}
\end{table}
The good algorithmic scaling of multigrid is demonstrated although the problems are noticeably harder at 32 and 64 planes, where the tolerance is reduced to $rtol=10^{-7}$, the refresh rate and V-cycle are modified, and a $V(6,6)$-cycle is used to compensate.

\subsection{Performance with standard user tolerance}
\label{sec:gpu09}

\begin{figure}[h!]
    \centering
    \includegraphics[width=0.6\linewidth]{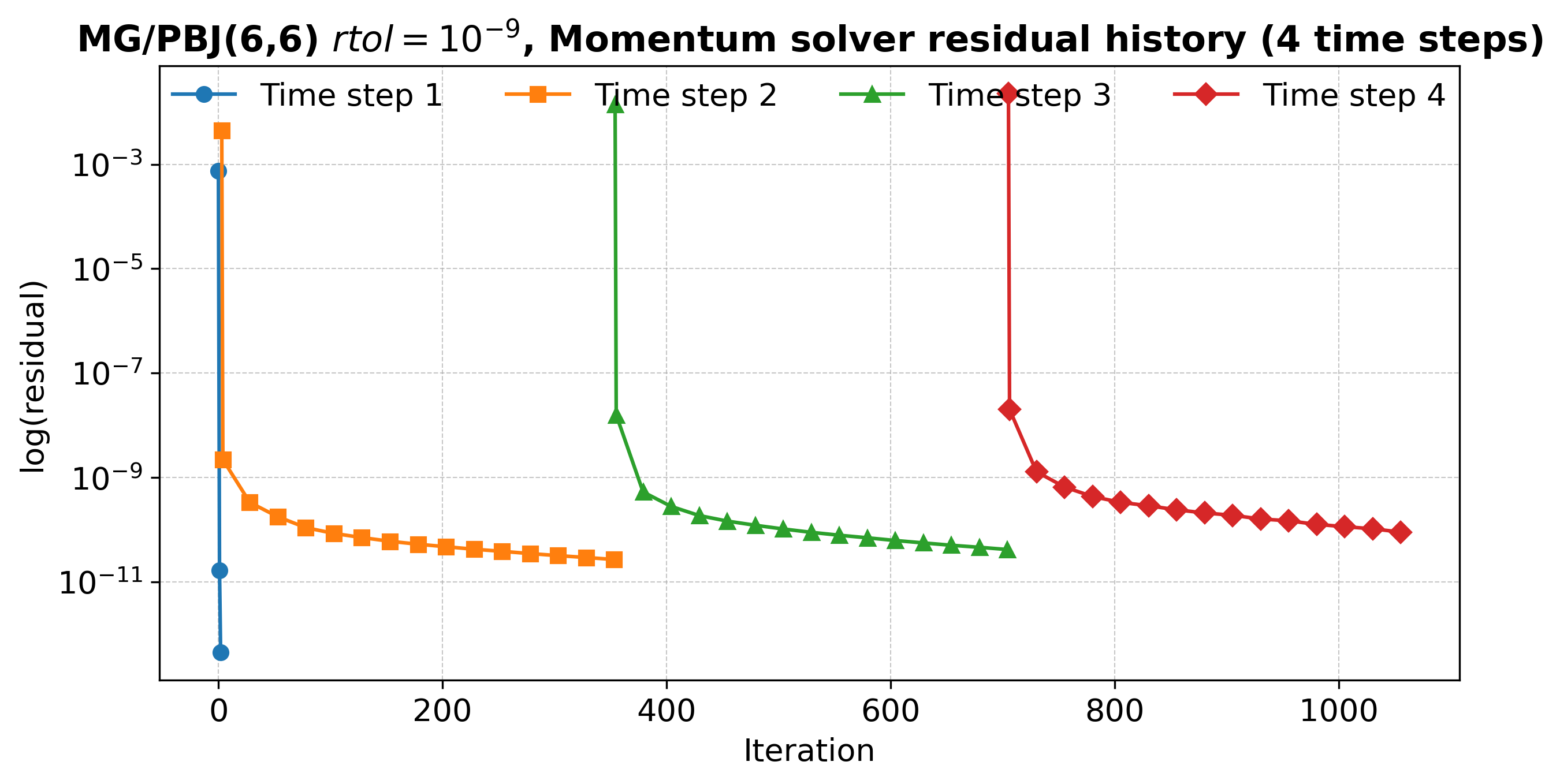}
    \caption{64 plane case with V(6,6) PBJ smoother. First three residuals plotted, every 25 thereafter.}
    \label{fig:stag1}
\end{figure}

The last section evaluated the CPU performance of the SPARC test problem with custom solver tolerances.
This section uses the high-accuracy tolerances used in practice on GPUs as well as CPUs.
We find that the PBJ solvers are not effective and the BJ solver needs to be used as the MG smoother.
The standard tolerance used in production is $rtol = 10^{-9}$ to provide a safe solver without the need for careful analysis for optimal parameters.
At this high level of accuracy, the PBJ smoother is prone to stagnation and is not viable (see Figure \ref{fig:stag1}).
To address this issue, the block Jacobi solver is re-purposed as the smoother and is used only once per iteration as the post-smoother.

\begin{figure}[h!]
    \centering
    \begin{tabular}{cc}
        \begin{subfigure}{0.5\textwidth}
            \centering
            \includegraphics[width=\linewidth]{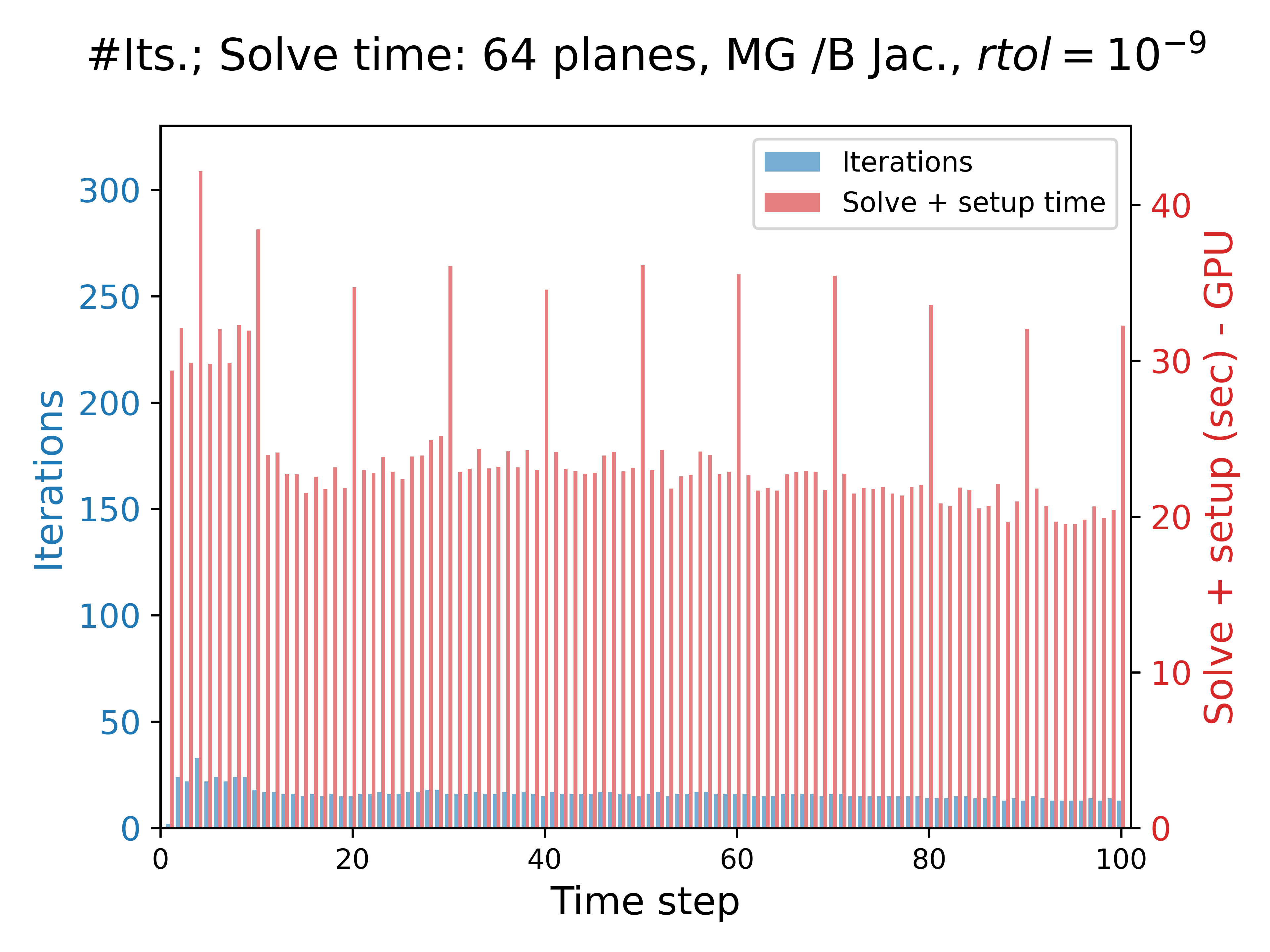}
            \caption{MG $V(0,1)$, block Jacobi smoother}
        \end{subfigure} & 
        \begin{subfigure}{0.5\textwidth}
            \centering
            \includegraphics[width=\linewidth]{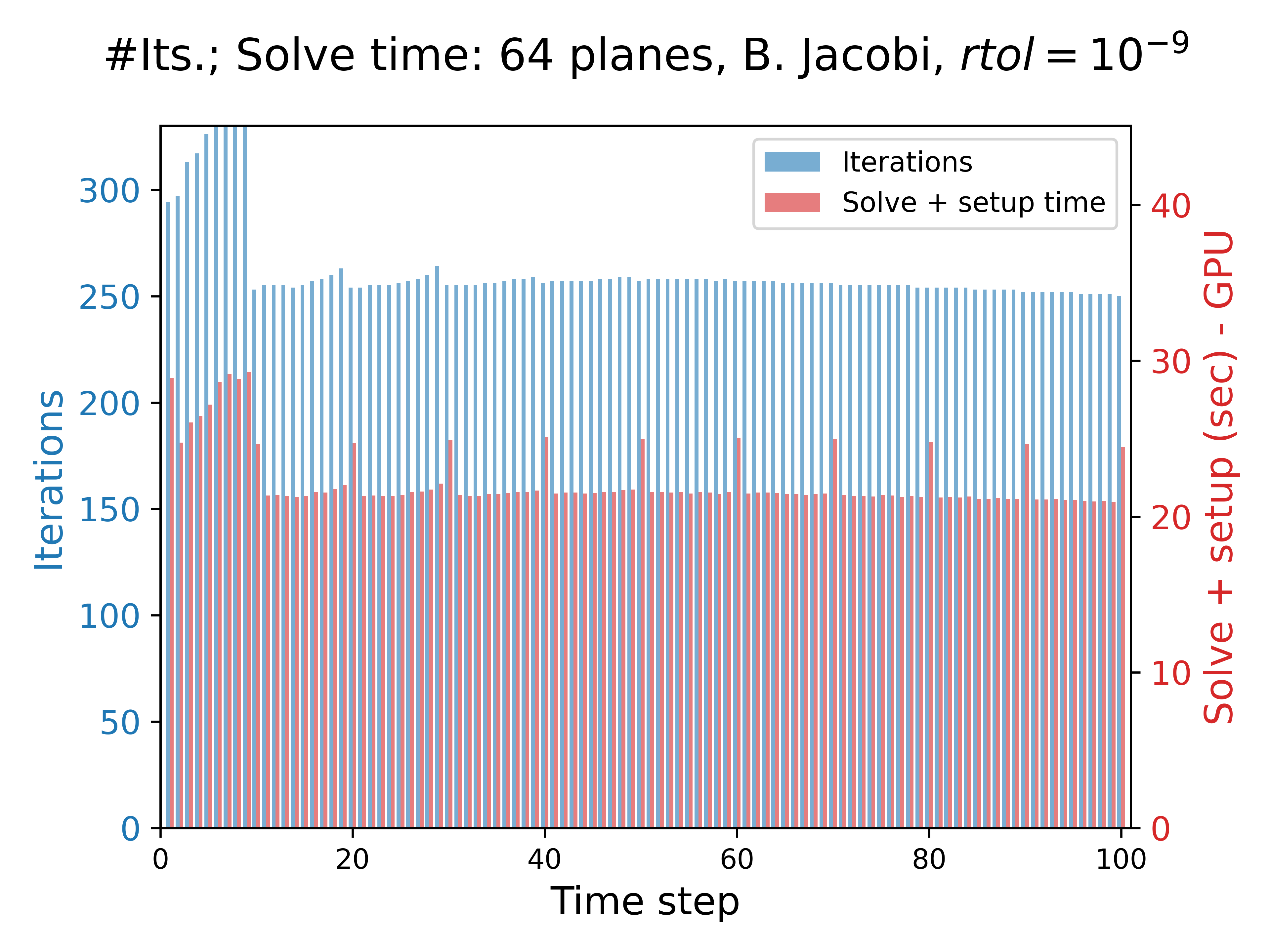}
            \caption{Block Jacobi solver}
        \end{subfigure} \\[1ex]
    \end{tabular}
  \caption{64 plane, Iterations (left axis/bar), GPU solve times (right axis/bar) vs time step. Solve times include periodic setup }
  \label{fig:64_pl_09}
\end{figure}

\begin{figure}[h!]
    \centering
    \begin{tabular}{cc}
        \begin{subfigure}{0.48\textwidth}
            \centering
            \includegraphics[width=\linewidth]{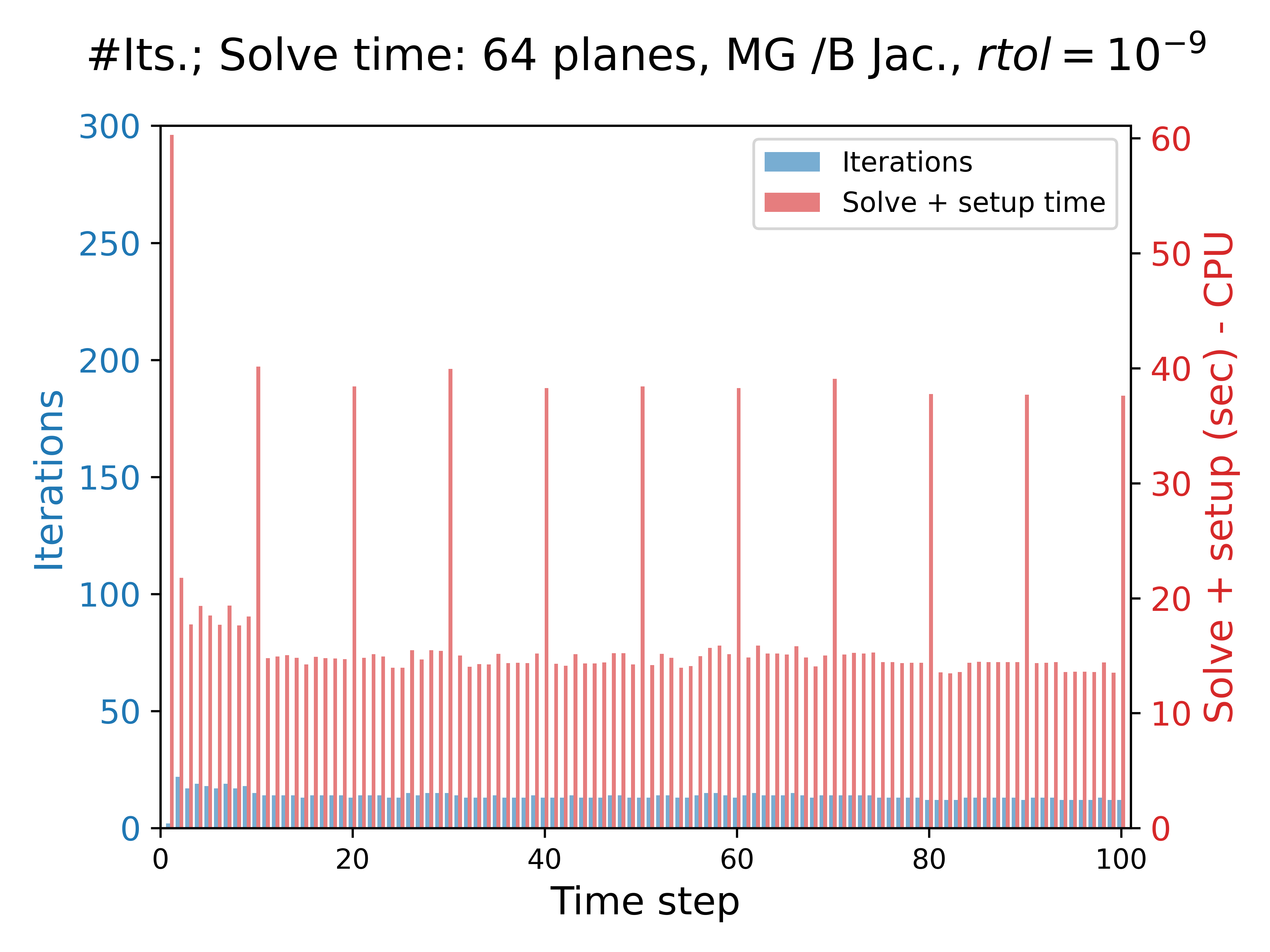}
            \caption{MG $V(0,1)$, block Jacobi smoother}
        \end{subfigure} & 
        \begin{subfigure}{0.48\textwidth}
            \centering
            \includegraphics[width=\linewidth]{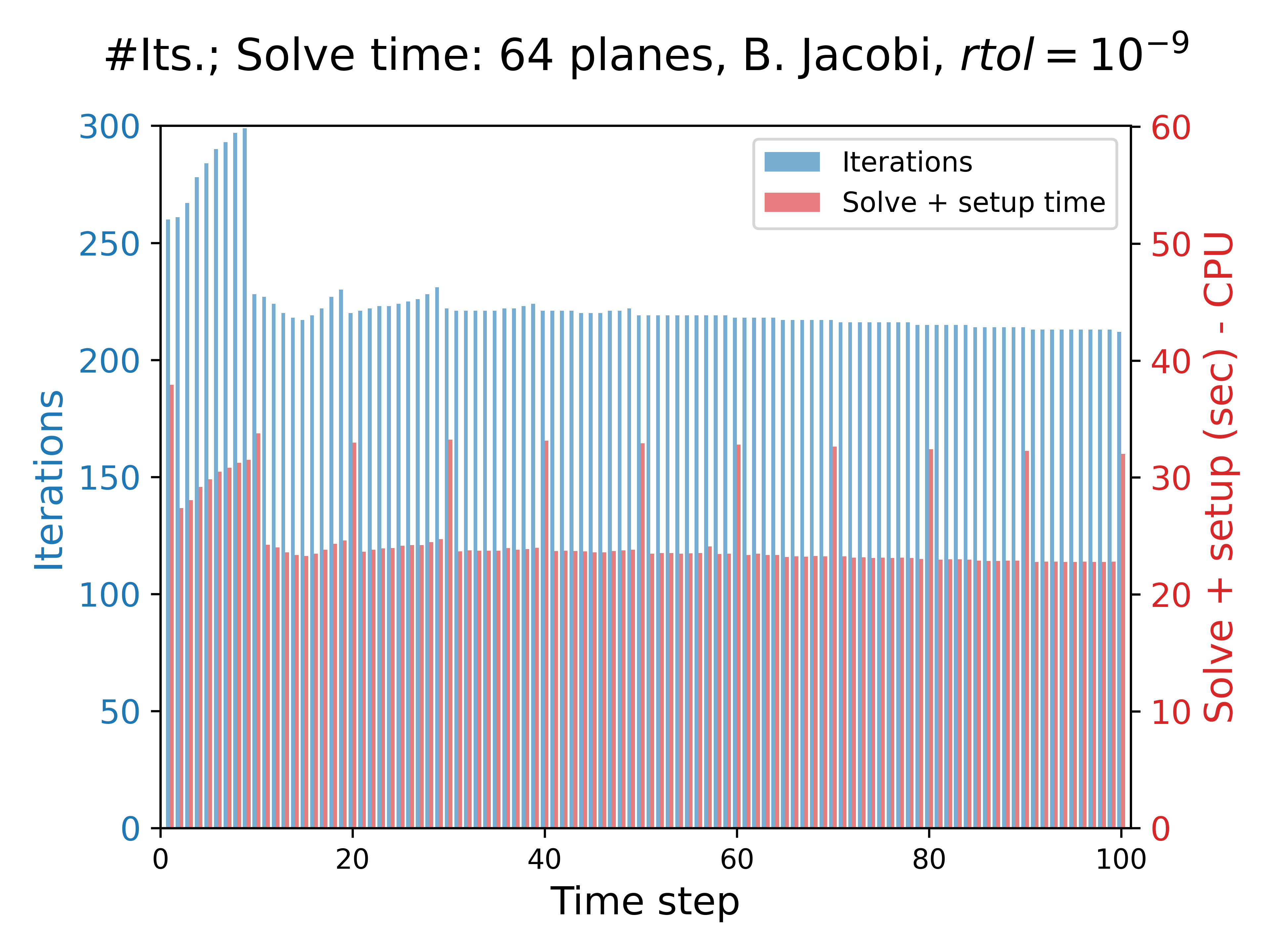}
            \caption{Block Jacobi solver}
        \end{subfigure} \\[1ex]
    \end{tabular}
  \caption{64 plane, Iterations (left axis/bar), CPU solve times (right axis/bar) vs time step. Solve times include periodic setup}
  \label{fig:64_pl_09_cpu}
\end{figure}

The Perlmutter GPU nodes have one AMD socket as opposed to two for the CPU nodes, which required the use of two GPU nodes, 8 NVIDIA A100 GPUs per plane, to keep the same number of processes and subdomain size per socket as in the CPU configuration.
Figure \ref{fig:64_pl_09} shows the solve time and iteration count history for the 64 plane case on GPUs with a solver tolerance of $rtol=10^{-9}$, and a solver refresh at every 10 time steps.
Figure \ref{fig:64_pl_09_cpu} shows data for this solve on the CPU configuration.

The solve phase performance, the time in the Krylov iterations after the setup, is similar for the CPU and GPU configurations in BJ, which can be attributed to the fact that MUMPS is run on the CPU and calls ScaLAPACK for dense LU solves in the multifrontal algorithm and is therefore not a full GPU solver.
The solve phase in LU has data dependencies that inhibit its performance on GPUs, whereas a sparse matrix vector product, the kernel of MG, has few data dependencies.
The solve time in the MG solver is significantly faster on the CPUs. 
We also observe some differences in the iteration counts in both solvers, which is not understood.
There are no known non-deterministic operations in this solver, but floating point differences could be the source of this issue -- the residual drops quickly in MG and the convergence rate, especially in the early time steps is very slow.
The condition number of these matrices is very high and the solver tolerance is very tight, which could explain these results.
The solve times for MG and BJ are similar on the GPU.
The iteration count with MG is reduced by about $16x$ from that of the BJ solver (Table \ref{tab:it_times_gpu}).

Table \ref{tab:it_times_gpu} shows the time in the PETSc solver, with setup costs, as measured with PETSc's ``KSPSolve" timer (with the total number of solver iterations over the 100 time steps) for multigrid $V(0,1)$ with the BJ smoother and the BJ solver with the GPU configuration. 
\begin{table}[h!]
    \centering
    \begin{tabular}{l|c}
    \hline
    \# planes      & 64\\
    \hline
    Multigrid V(0,1) / BJ  &  2,150 (1,653) \\ 
    block Jacobi (BJ)  &  2,188 (26,203) \\
    \hline
    \end{tabular}
    \caption{Total solver time including solver setup (total iterations) in seconds on GPU nodes, $rtol = 10^{-9}$}
    \label{tab:it_times_gpu}
\end{table}

\textit{Complexity analysis}: This data shows that the total solve time is about the same with the multigrid solver and the total iteration count, and hence number of block Jacobi applications on the fine grid, is reduced by a factor of about $16x$.
With a $V(0,1)$-cycle, MG applies the block Jacobi solver once per iteration on the fine grid, as well as once on every coarse grid.
The work and memory complexity in terms of LU solves is a geometric sum $1 + \frac{1}{2}+\frac{1}{4}+ \frac{1}{8} + ...$ of that of the fine grid, which results in an asymptotic work complexity of $2x$ the BJ solver. 
This is a lower bound on the cost of the MG solver.
The computational depth with seven levels is $7x$ higher in the MG solver, assuming the cost of every grid is the same, which indicates an upper bound of $7x$ more cost per iteration with MG / BJ smoothing.
The $16x$ fewer iterations with MG and approximately equal solve times suggests the MG times are about $2x$ higher than the upper bound with this simple model.
This indicates that coarse grids are performing poorly, which is the subject of future work.

\subsubsection{Setup Cost}

The refresh costs are not measured separately in the \textit{M3D-C1} timers, but they can be inferred from the jump in the solve time at the refresh period.
The cost per iteration can be assumed to be constant.
The MG refresh costs are higher than the BJ solver from the Galerkin coarse grid construction and the first setup is larger due to symbolic setup of this Galerkin operator.
The MG solver LU factorization costs are higher than the BJ solver because the $V(0,1)$-cycle factors matrices on seven levels instead of one.
Table \ref{tab:petsc-setup-timing} shows the cost of the setup inferred from Figures \ref{fig:64_pl_09} and \ref{fig:64_pl_09_cpu}.
\begin{table}[ht]
  \centering
  \caption{Setup cost at refresh periods in seconds (cost of symbolic setup in first time step)}
  \label{tab:petsc-setup-timing}
  \begin{tabular}{lrr}
    \hline
     & GPU & CPU \\
    \hline
    MG $V(0,1)$-cycle & 12 (14) & 25 (33) \\
    BJ                & 3 (1) & 9 (1) \\
    \hline
  \end{tabular}
\end{table}
This data shows a significant speedup with GPUs in the setup phase.
This can be attributed to the high arithmetic intensity (flops per byte of data) of both the matrix triple product in the Galerkin coarse grid construction and the LU factorizations.


\subsection{\label{subsec:stell} Stellarator problem performance}

The stellarator simulations are performed on a semi-structured grid of 8 toroidal planes with $1.88 \times 10^5$ three-dimensional elements.  
The stellarator test described in~\S\ref{sec:stell} uses the production solver tolerance $\mathit{rtol} = 10^{-9}$.
This test is relatively small with 8 toroidal planes with $1.88 \times 10^5$ three-dimensional elements and about $3.4 \times 10^6$ equations.
The focus in this experiment is on the convergence rate of the solvers.
Unlike the SPARC problem, the block Jacobi solver fails to converge, while the multigrid solver with BJ smoothing converges, demonstrating a robustness advantage of MG methods on non-axisymmetric configurations.
Note, this project is under active development and these results were generated in early 2026.

\begin{figure}[h!]
    \centering
    \begin{tabular}{cc}
        \begin{subfigure}{0.48\textwidth}
            \centering
            \includegraphics[width=\linewidth]{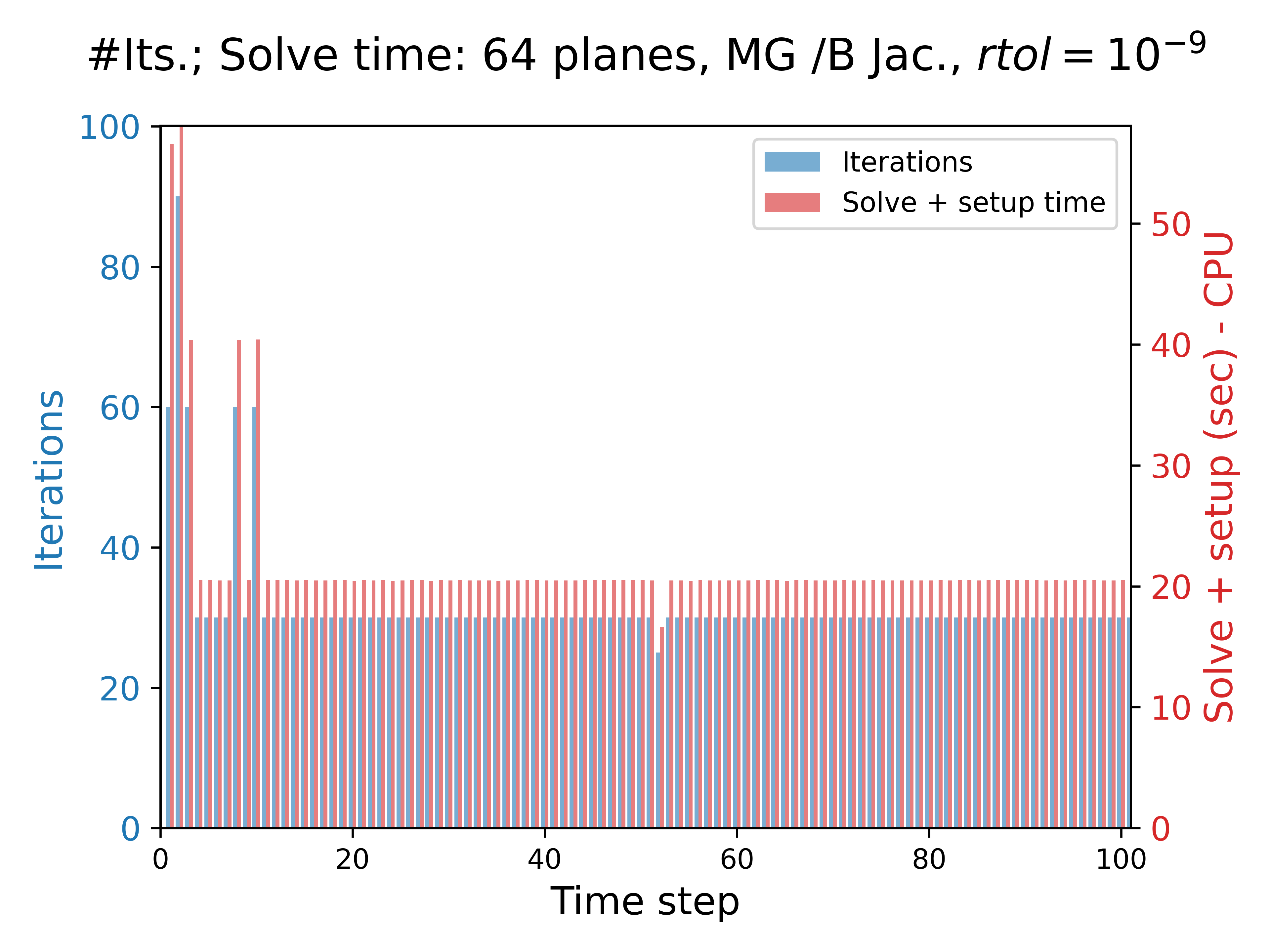}
            \caption{MG $V(0,1)$, block Jacobi smoother}
        \end{subfigure} & 
        \begin{subfigure}{0.48\textwidth}
            \centering
            \includegraphics[width=\linewidth]{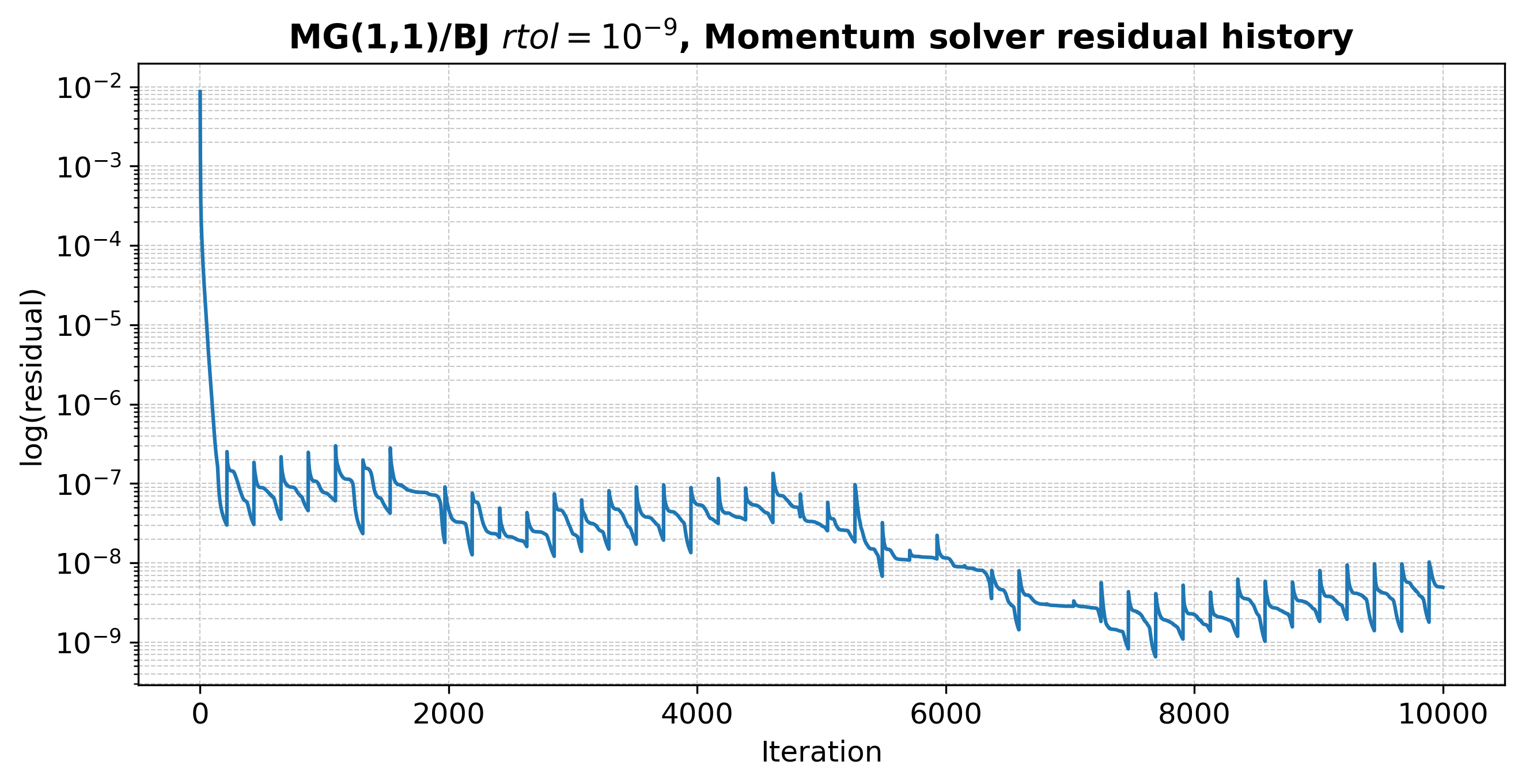}
            \caption{Block Jacobi solver residual history of failed solve}
        \end{subfigure} \\[1ex]
    \end{tabular}
  \caption{Stellarator 8 plane case: Iterations and solve times vs time step (left), stagnated residual history (right) using the BJ solver (jumps result from FGMRES restarts)}
  \label{fig:stell_09}
\end{figure}

Figure \ref{fig:stell_09} (left) shows the MG iteration count history for the stellarator test with $rtol=10^{-9}$, and a solver refresh at every time step.
The block Jacobi solver is used as the smoother once per iteration as the post-smoother and the pre-smoother is one iteration of PBJ, which we call a $V(0,1)$-cycle for clarity.
Figure \ref{fig:stell_09} (right) shows the residual history using the failed BJ solve.
This data shows that the MG solver is converging on par with the SPARC test, while the BJ solver stagnates.

\section{\label{sec:future} Conclusion and future work}


This work develops a toroidal semi-coarsening, geometric multigrid (MG) solver, with point-block Jacobi (PBJ) and full plane block Jacobi (BJ) smoothers, and algebraic coarse grids, in \textit{M3D-C1}.
The new MG solver is competitive on the largest version of one test case (SPARC), much faster on some smaller configurations, and is effective on a stellarator test where the BJ solver fails.

Several optimizations remain to be explored. 
The most urgent issue is the coarse grid data model where redundant coarse grid solves were required for the SPARC tests due to direct solver MPI failures on coarse grids during the factorization stage with both \textit{SuperLU} and \textit{MUMPS}.
A better approach to reducing the degree of parallelism, to relieve pressure on the LU solvers, is to leave coarse grid processors idle as is often done in scalable MG solvers \cite{Adams04}.
Additionally, redundant coarse grid solvers significantly complicate the solver parameters (Appendix \ref{sec:params}).

The most impactful optimization, and the most complex to implement, is adding $2D$ multigrid to the poloidal plane, resulting in a full $3D$ multigrid solver, which is a complex project requiring unstructured geometric coarse grids \cite{Adams04}.
$3D$ MG will reduce the work and memory complexity of the solver substantially in coarse grid solves and the coarse grid operator construction.

Another area of improvement for this solver is expanding the options for, and size of, the blocks in the PBJ smoothers that fit in a device's thread group (e.g., an NVIDIA SM or \textit{cooperative group}).
The blocks in the current PBJ smoothers can be enlarged to some extent, so as to provide more powerful smoothers, before the cost of the dense direct solves becomes prohibitive.
The appeal of PBJ smoothers is the ability to use dense linear algebra (explicit inverses and BLAS2 solves), which performs exceptionally well on GPUs.
Increasing the block size adds sparsity to these sub-solves that would need to be padded, limiting the size.
For example, a $3^3$ vertex subdomain would have 972 equations with a small amount of padding and would be tractable on existing GPU architectures.
Sparse batch solvers are a potential option for larger subdomains.
PETSc does have sparse batched Krylov solvers with Jacobi preconditioners \cite{adams2024performance}, and algebraic MG batched preconditioners have been added to PETSc recently.

The main choices for enlarging these blocks are 1) line subdomains (e.g., in the toroidal direction), 2) plane subdomains (e.g., in the poloidal plane), or 3) isotropic subdomains.
It is possible that different subdomains would be optimal for each of the three equations ($U, \omega, \chi$) in a $3 \times 3$ block outer solver -- a \textit{FieldSplit} solver in PETSc -- because each of these equations represents a distinct wave in the system and will invariably have different anisotropic characteristics.
This splitting of waves also nearly decouples the three equations (Appendix \ref{sec:blocknorms}), which is required for the (3) block Gauss-Seidel iteration in a \textit{FieldSplit} solver to converge well.
More effective \textit{FieldSplit} solvers (i.e., Schur complement based methods) are not tractable for the \textit{M3D-C1} matrices.

\textit{Reproducibility} is discussed in Appendix \ref{sec:params}.

\section*{Acknowledgments}

This work was supported in part by the U.S. Department of Energy,
Office of Science, Office of Fusion Energy Sciences and Office of Advanced Scientific Computing Research,
Scientific Discovery through Advanced Computing (SciDAC) program
through the FASTMath Institute under Contract No. DE-AC02-05CH11231 at
Lawrence Berkeley National Laboratory.
This research used resources of the National Energy Research
Scientific Computing Center, a DOE Office of Science User Facility
supported by the Office of Science of the U.S. Department of Energy
under Contract No. DE-AC02-05CH11231 using NERSC award
ASCR-ERCAP0030112.
This manuscript has been [co-]authored by Princeton University/Princeton Plasma Physics Laboratory under contract number DE-AC02-09CH11466 with the U.S.  Department of Energy. The United States Government retains a non-exclusive, paid-up, irrevocable, world-wide license to publish or reproduce the published form of this manuscript, or allow others to do so, for United States Government purpose(s).
DOE SciDAC program, Center for Edge of Tokamak OPtimization (CETOP), under Award Number DE-AC02-09CH11466.

\bibliographystyle{plain}
\bibliography{the,doc}

\appendix

\section{The resistive MHD equations}
\label{sec:appendix}

This appendix sketches the numerical methods in the evolution of the momentum equations (\ref{eq:mom1}) with a focus on the algebraic form of the diagonal blocks of the Jacobian matrix in \textit{M3D-C1}, which is of the most relevance to algebraic solvers. 
Magnetohydrodynamics (MHD) treats a magnetized plasma as a conducting fluid with mean number density $n$, ion mass $M_i$, velocity $\v$ and current $\J$, Maxwell's equations and conservation laws (velocity moments $0-2$), of the form:

Start from the resistive MHD equations:

\begin{eqnarray}
	\ddt{n} + \div{(n \v)} =0 & \mbox{continuity}
    \\
	\ddt{\B} = \curl{ \left[ \v \times \B -\eta \J \right] } & \mbox{Faraday's law}
    \\
	nM_i \frac{d\v}{dt} + \grad{p} - \nu \nabla^2 \v = \J \times \B & \mbox{momentum} \label{eq:mom1}
	\\
	\E + \v \times \B = \eta \J + \frac{1}{ne} \left( \J \times \B - \grad{p}_e - \div{\P}_e \right) & \mbox{Ohm's law}
	\\
	\ddt{p} + \v \cdot \grad{p} +\gamma p \div{\v}   = (\gamma -1) \left[ \eta\J^2 - \P:\grad{\v} \right]  & \mbox{energy in conservation form}
\end{eqnarray}
\noindent
with electrical resistivity $\eta$, viscosity $\mu$ and a pressure tensor $\P$.
\textit{M3D-C1} is an extended MHD, two-fluid, model with electron charge $e$, an electron pressure tensor ${\P}_e$ and electron pressure $p_e$.
The total pressure $p$ and adiabatic index $\gamma$ are from the equation of state for internal energy.

\subsection{Dimensionless scaling of the momentum equation}

These MHD equations are scaled by physically relevant quantities, which regularizes the numerics, simplifies the notation, and results in units that are physically relevant.

\begin{eqnarray}
	nM_i \frac{d\v}{dt} + \grad{p} - \nu \nabla^2 \v = \J \times \B \\
	n_0\tilde{n}M_i \frac{v_0}{t_0}\frac{d\tilde{\v}}{d\tilde{t}} 
	+ \frac{p_0}{l_0} \tilde{\nabla}\tilde{p} 
	- \frac{\nu_0 v_0}{l_0^2}\tilde{\nu} \tilde{\nabla}^2 \tilde{\v} = J_0B_0 \tilde{\J} \times \tilde{\B} \\
	\frac{l_0^2}{t_0^2} \frac{\mu_0 n_0 M_i}{B_0^2} \tilde{n} \frac{d\tilde{\v}}{d\tilde{t}} 
	+ \frac{\mu_0 p_0}{B_0^2} \tilde{\nabla}\tilde{p} -\frac{\nu_0 v_0 \mu_0}{B_0^2 l_0}\tilde{\nu} \tilde{\nabla}^2 \tilde{\v} = \tilde{\J} \times \tilde{\B} \\
	\tilde{n} \frac{d\tilde{\v}}{d\tilde{t}} + \tilde{\nabla}\tilde{p} -\tilde{\nu} \tilde{\nabla}^2 \tilde{\v} = \tilde{\J} \times \tilde{\B} 
\end{eqnarray}
where 
$$
t_0 = l_0 \left[\frac{\mu_0 n_0 M_i}{B_0^2} \right]^{\frac{1}{2}} ,
p_0 = \frac{B_0^2}{\mu_0},
J_0 = \frac{B_0}{\mu_0 l_0},
\v_0 = \frac{l_0}{t_0},
\nu_0 = \frac{B_0^2 t_0}{\mu_0} = \frac{n_0 M_i}{t_0} l_0^2
$$

\subsection{Parabolization}
\label{sec:parb}

Parabolization is critical in forming equations that are elliptic and in \textit{M3D-C1} symmetric.
This parabolization process results in a fourth order curl operator (\ref{eq:L1}) and is included here for reference.

Let $\dot{\v} \equiv \ddt{\v}$ and the ideal MHD momentum equation results:
\begin{equation}
n_0 \dot{\v} + \grad{p} = \left[ \left( \nabla \times \B \right) \times \B \right]
\end{equation}
take its time derivative 
\begin{equation}
	n_0 \ddot{\v} + \grad{\dot{p}} = \left[ \left( \nabla \times \dot{\B} \right) \times \B + 
								  \left( \nabla \times \B \right) \times \dot{\B}
	                                                  \right]
\end{equation}
Substitute the ideal MHD equation for $\dot{\B}$ and $\dot{p}$
\begin{eqnarray}
	\dot{\B} = \nabla \times \left[ \v \times \B \right]\\
	\dot{p} = -\v \cdot \grad{p} - \gamma p \div{\v}
\end{eqnarray}
we get the parabolization operator
\begin{equation}
	\begin{array}{l}
\mathcal{L} (\v) = \left\{ \nabla \times \left[ \nabla \times \left( \v \times \B \right) \right] \right\} \times \B 
 +  \left\{ \left( \nabla \times \B \right) \times \nabla \times \left( \v \times \B \right) \right\} 
 + \grad{ ( \v \cdot \grad{p} + \gamma p \div{\v} ) }
\end{array}
\label{eq:L1}
\end{equation}

with the parabolization, the momentum equation becomes:
\begin{eqnarray}
	\left\{ n - \theta^2 \dt^2 \mathcal{L} \right\} \pp{\v}{t} + n \v \cdot \grad{\v} + \grad{p} - \nu \nabla^2 \v 
	= \left[ \left( \nabla \times \B \right) \times \B \right] 
    \label{eq:L2}
\end{eqnarray}
Then we can advance the velocity as follows
\begin{equation}
	\begin{array}{l}
\left\{ n - \theta^2 \dt^2 \mathcal{L} \right\} \v^{n+1} + \theta \dt (n \v^{n+1} \cdot \grad{\v^n} + n \v^n \cdot \grad{\v^{n+1}}) -\nu \theta \dt \nabla^2 \v^{n+1} = \\
\qquad \qquad \qquad \qquad \qquad \qquad
\left\{ n - \theta^2 \dt^2 \mathcal{L} \right\} \v^{n} -(1-2\theta) \dt (n \v^n \cdot \grad{\v^n}) +
\dt \left\{ - \grad{p} + \left[ \left( \nabla \times \B \right) \times \B \right] \right\}^{n+\frac{1}{2}}
\end{array}
\label{eq:L3}
\end{equation}

\subsection{Decompose the velocity in \texorpdfstring{$(R,\phi,Z)$}{R,phi,Z} coordinate system}
\label{sec:momentum}

An important feature of the \textit{M3D-C1} formulation is the decomposition of velocity into three scalar fields that expand to MHD waves: $U,\omega, \chi$

\begin{eqnarray}
	\v = R^2 \grad{U} \times \grad{\phi} + R^2 \omega \grad{\phi} + \frac{1}{R^2} \nabla_\perp \chi
  \label{eq:stream}  
\end{eqnarray}

\noindent
The operator $\nabla_\perp$ denotes the gradient in the $(R,Z)$ plane, 
$$
\nabla_\perp \chi \equiv \chi_R \hat{R} + \chi_Z \hat{Z},
$$
orthogonal to $\grad{\phi}$. $| \grad{\phi} | = \ff{1}{R}$.

$U$ - shear Alfven wave, does not compress the toroidal magnetic field.

$\omega$ - slow wave, does not compress the toroidal magnetic field.

$\chi$ - fast wave, does compress the toroidal field. 

These equations are meant to give a sense of the nature of the equations being solved in the momentum solve. 

\section{Reproducibility issues, \textit{M3D-C1} and PETSc options}
\label{sec:params}

\textit{M3D-C1} is open-source \url{https://github.com/PrincetonUniversity/M3DC1}, but this section is intended to aid in producing the results in this paper by \textit{M3D-C1} users.
The complete input instructions would not be feasible to include here and one would need to work with the \textit{M3D-C1} team to set this problem up.
This input deck is what needs to be provided by the solver team, the authors, and one would need to work with the physics team to reproduce these results.
Details of the physical model are published by Datta \cite{Datta2025}.
The data, run and plot scripts, inputs and README-solver-paper.txt file for reproducibility data are in {\tt ...mp288/jinchen/M3DC1/WORK/SPRC-m04-3D-03-TEST} at NERSC.

This is an example of the PETSc input that one could provide as the \texttt{-options\_file} argument to \textit{M3D-C1}, or place in a .petscrc file.
This file is for a 64 plane case.
The {\tt -pc\_bjacobi\_blocks 64} option must be set to the number of planes and the degree of redundancy is a tunable parameter.
The block size must be set for each level (e.g., \texttt{-hard\_mg\_levels\_1\_up\_redundant\_pc\_bjacobi\_blocks 2}), and one can run with say 32 planes and the \texttt{-hard\_mg\_levels\_6\_up\_pc\_bjacobi\_blocks 64} line will be safely ignored, thus only {\tt -pc\_bjacobi\_blocks 64} need be changed for fewer planes.
\begin{Verbatim}[fontsize=\small]
-pc_bjacobi_blocks 64 # set to number of planes
-hard_ksp_atol 1.e-20 # (source: file)
-hard_ksp_converged_reason # (source: file)
-hard_ksp_gmres_preallocate # (source: file)
-hard_ksp_gmres_restart 200 # (source: file)
-hard_ksp_max_it 200 # (source: file)
-hard_ksp_min_it 1 # (source: command line)
-hard_ksp_monitor # (source: file)
-hard_ksp_rtol 1.e-16 # (source: command line)
-hard_ksp_type fgmres # (source: file)
-hard_ksp_view # (source: command line)
-hard_mg_coarse_ksp_type preonly # (source: file)
-hard_mg_coarse_pc_redundant_number 16 # (source: file)
-hard_mg_coarse_pc_type redundant # (source: file)
-hard_mg_coarse_redundant_ksp_converged_reason # (source: file)
-hard_mg_coarse_redundant_ksp_rtol 1.e-12 # (source: file)
-hard_mg_coarse_redundant_ksp_type gmres # (source: file)
-hard_mg_coarse_redundant_pc_factor_mat_solver_type mumps # (source: file)
-hard_mg_coarse_redundant_pc_type lu # (source: file)
-hard_mg_levels_1_up_ksp_max_it 1 # (source: file)
-hard_mg_levels_1_up_ksp_norm_type none # (source: file)
-hard_mg_levels_1_up_ksp_type gmres # (source: file)
-hard_mg_levels_1_up_pc_redundant_number 8 # (source: file)
-hard_mg_levels_1_up_pc_type redundant # (source: file)
-hard_mg_levels_1_up_redundant_ksp_type preonly # (source: file)
-hard_mg_levels_1_up_redundant_pc_bjacobi_blocks 2 # (source: file)
-hard_mg_levels_1_up_redundant_pc_type bjacobi # (source: file)
-hard_mg_levels_1_up_redundant_sub_ksp_type preonly # (source: file)
-hard_mg_levels_1_up_redundant_sub_pc_factor_mat_solver_type mumps # (source: file)
-hard_mg_levels_1_up_redundant_sub_pc_type lu # (source: file)
-hard_mg_levels_2_up_ksp_max_it 1 # (source: file)
-hard_mg_levels_2_up_ksp_norm_type none # (source: file)
-hard_mg_levels_2_up_ksp_type gmres # (source: file)
-hard_mg_levels_2_up_pc_redundant_number 4 # (source: file)
-hard_mg_levels_2_up_pc_type redundant # (source: file)
-hard_mg_levels_2_up_redundant_ksp_type preonly # (source: file)
-hard_mg_levels_2_up_redundant_pc_bjacobi_blocks 4 # (source: file)
-hard_mg_levels_2_up_redundant_pc_type bjacobi # (source: file)
-hard_mg_levels_2_up_redundant_sub_ksp_type preonly # (source: file)
-hard_mg_levels_2_up_redundant_sub_pc_factor_mat_solver_type mumps # (source: file)
-hard_mg_levels_2_up_redundant_sub_pc_type lu # (source: file)
-hard_mg_levels_3_up_ksp_max_it 1 # (source: file)
-hard_mg_levels_3_up_ksp_norm_type none # (source: file)
-hard_mg_levels_3_up_ksp_type gmres # (source: file)
-hard_mg_levels_3_up_pc_redundant_number 2 # (source: file)
-hard_mg_levels_3_up_pc_type redundant # (source: file)
-hard_mg_levels_3_up_redundant_ksp_type preonly # (source: file)
-hard_mg_levels_3_up_redundant_pc_bjacobi_blocks 8 # (source: file)
-hard_mg_levels_3_up_redundant_pc_type bjacobi # (source: file)
-hard_mg_levels_3_up_redundant_sub_ksp_type preonly # (source: file)
-hard_mg_levels_3_up_redundant_sub_pc_factor_mat_solver_type mumps # (source: file)
-hard_mg_levels_3_up_redundant_sub_pc_type lu # (source: file)
-hard_mg_levels_4_up_pc_bjacobi_blocks 16 # (source: file)
-hard_mg_levels_5_up_pc_bjacobi_blocks 32 # (source: file)
-hard_mg_levels_6_up_pc_bjacobi_blocks 64 # (source: file)
-hard_mg_levels_ksp_max_it 1 # (source: file)
-hard_mg_levels_ksp_norm_type none # (source: file)
-hard_mg_levels_ksp_type gmres # (source: file)
-hard_mg_levels_pc_type pbjacobi # (source: file)
-hard_mg_levels_up_ksp_converged_reason # (source: file)
-hard_mg_levels_up_ksp_gmres_restart 1 # (source: file)
-hard_mg_levels_up_ksp_max_it 1 # (source: file)
-hard_mg_levels_up_ksp_norm_type none # (source: file)
-hard_mg_levels_up_ksp_type gmres # (source: file)
-hard_mg_levels_up_pc_type bjacobi # (source: file)
-hard_mg_levels_up_sub_ksp_type preonly # (source: file)
-hard_mg_levels_up_sub_pc_factor_mat_solver_type mumps # (source: file)
-hard_mg_levels_up_sub_pc_type lu # (source: file)
-hard_pc_mg_distinct_smoothup # (source: file)
-ihard_mat_type aijkokkos # (source: command line)
-ksp_atol 1.e-20 # (source: file)
-ksp_converged_reason # (source: file)
-ksp_error_if_not_converged # (source: file)
-ksp_max_it 100 # (source: file)
-ksp_monitor_5 # (source: command line)
-ksp_rtol 1.e-9 # (source: file)
-ksp_type gmres # (source: file)
-log_view :log_mg16_bj.txt # (source: command line)
-mgfs 5 # (source: command line)
-mhard_mat_type aijkokkos # (source: command line)
-on_error_abort # (source: file)
-options_left # (source: file)
-pc_type bjacobi # (source: file)
-sub_ksp_type preonly # (source: file)
-sub_pc_factor_mat_solver_type mumps # (source: file)
-sub_pc_type lu # (source: file)


\end{Verbatim}

\section{Block norms}
\label{sec:blocknorms}

This section reports the norms of these blocks, a $3 \times 3$ scalar matrix, and the degree of diagonal dominance is indicative of the convergence rates that one can expect with a \textit{FieldSplit} solver, a block Gauss-Seidel iteration, and illustrates the near diagonalization that results from the velocity decomposition used in the momentum solver.
The relative off-diagonal block norms range between $10^{-2}$ and $10^{-3}$. 

Note: these block norm tests are in the context of plane solvers. 
We first extracted a single block from a full 3D matrix, on which the analysis was performed. 

To get a sense of the strength of coupling between fields, we consider symmetric diagonal scaling of the Frobenius norms of each field block. Specifically, we examine the following 3$\times$3 coupling matrix for three test cases.

\[
\begin{pmatrix}
1 & \frac{|| A_{U,\omega} ||_F}{\sqrt{|| A_{U,U} ||_F || A_{\omega,\omega} ||_F}} & \frac{|| A_{U,\chi} ||_F}{\sqrt{|| A_{U,U} ||_F || A_{\chi,\chi} ||_F}} \\
\frac{|| A_{\omega,U} ||_F}{\sqrt{|| A_{U,U} ||_F || A_{\omega,\omega} ||_F}} & 1 & \frac{|| A_{\omega,\chi} ||_F}{\sqrt{|| A_{\omega,\omega} ||_F || A_{\chi,\chi} ||_F}} \\
\frac{|| A_{\chi,U} ||_F}{\sqrt{|| A_{U,U} ||_F || A_{\chi,\chi} ||_F}} & \frac{|| A_{\chi,\omega} ||_F}{\sqrt{|| A_{\omega,\omega} ||_F || A_{\chi,\chi} ||_F}} & 1
\end{pmatrix}
\]

The number of equations and number of non-zeros for the three test cases are given in Table  \ref{tab:test_cases}. 

\begin{table}[ht]
\centering
\begin{tabular}{|c|c|c|c|c|}
\hline
 & Case 1 & Case 2 & Case 3 & Case RE (model problem)\\
\hline
Test case name & p-cpu-16F-R01 & nstx\_120446 & numvar.NV3.mgfs & SPARC - disruption \\
\hline
N & 189 180 & 365 796 & 401 292 & 1,879,920\\
\hline
NNZ & 46 472 313 & 89 557 184 & 98 771 994 & 1,409,524,416\\
\hline
\end{tabular}
\caption{Number of equations and non-zeros in three test cases from the \textit{M3D-C1} test case library}
\label{tab:test_cases}
\end{table}

\noindent
The coupling matrices for each case, RE being the model problem in this work, are:

\[ \text{Case 1: }\left( 
\begin{array}{ccc}
1.0 & 1.5 \times 10^{-4} & 1.3 \times 10^{-3} \\ 
1.5 \times 10^{-4}  & 1.0 & 8.5 \times 10^{-4} \\
1.3 \times 10^{-3}  & 8.5 \times 10^{-4} & 1.0 \\
\end{array} \right)
\qquad\quad
\text{Case 2: } \left( \begin{array}{ccc}
1.0 & 2.3 \times 10^{-4} & 8.4 \times 10^{-3} \\ 
2.3 \times 10^{-4}  & 1.0 & 1.3 \times 10^{-3} \\
8.4 \times 10^{-3}  & 1.3 \times 10^{-3} & 1.0 \\
\end{array}\right)
\]

\[ \text{Case 3: }\left( 
\begin{array}{ccc}
1.0 & 5.1 \times 10^{-5} & 6.3 \times 10^{-3} \\ 
5.1 \times 10^{-5}  & 1.0 & 1.6 \times 10^{-2} \\
6.3 \times 10^{-3}  & 1.6 \times 10^{-2} & 1.0
\end{array} \right)
\qquad\quad
\text{Case RE: } \left( \begin{array}{ccc}
1.0 & 4.6 \times 10^{-4} & 2.9 \times 10^{-2} \\ 
4.6 \times 10^{-4}  & 1.0 & 9.7 \times 10^{-3} \\
2.9 \times 10^{-2}  & 9.7 \times 10^{-3} & 1.0 \\
\end{array}\right)
\]

\end{document}